%% file: preprint.tex
\documentclass{article}

\usepackage{multirow}
\usepackage{subcaption}
\usepackage{authblk}

\usepackage{amssymb}

\usepackage[numbers,sort&compress]{natbib}
\usepackage[colorlinks=true,colorlinks,
linkcolor=purple,
citecolor=purple,
urlcolor=blue,
filecolor=blue]{hyperref}
\usepackage[noabbrev,capitalise,nameinlink]{cleveref}

\usepackage{float}
\usepackage[all]{nowidow}

\usepackage{url}

\usepackage{tikz}  % for circled command
\newcommand*\circled[1]{\tikz[baseline=(char.base)]{
\node[shape=circle,draw,inner sep=0.75pt,scale=0.8,font=\fontfamily{cmss}\selectfont] (char) {#1};}%
}

\definecolor{customblue}{HTML}{4477AA}
\definecolor{customred}{HTML}{EE6677}
\definecolor{customgreen}{HTML}{228833}
\newcommand*{\textblue}[1][\textcolor{customblue}]{#1}
\newcommand*{\textred}[1][\textcolor{customred}]{#1}
\newcommand*{\textgreen}[1][\textcolor{customgreen}]{#1}

\begin{document}

\title{Managing Data Replication and Distribution in the Fog with FReD}

\author{Tobias Pfandzelter}
\author{Nils Japke}
\author{Trever Schirmer}
\author{Jonathan Hasenburg}
\author{David Bermbach}
\affil{\textit{Technische Universit\"at Berlin \& Einstein Center Digital Future}\\
    \textit{Mobile Cloud Computing Group}\\
    \texttt{\{tp,nj,ts,jh,db\}@mcc.tu-berlin.de}}

\date{}

\maketitle

\begin{abstract}
    The heterogeneous, geographically distributed infrastructure of fog computing poses challenges in data replication, data distribution, and data mobility for fog applications.
    Fog computing is still missing the necessary abstractions to manage application data, and fog application developers need to re-implement data management for every new piece of software.
    Proposed solutions are limited to certain application domains, such as the IoT, are not flexible in regard to network topology, or do not provide the means for applications to control the movement of their data.

    In this paper, we present FReD, a data replication middleware for the fog.
    FReD serves as a building block for configurable fog data distribution and enables low-latency, high-bandwidth, and privacy-sensitive applications.
    FReD is a common data access interface across heterogeneous infrastructure and network topologies, provides transparent and controllable data distribution, and can be integrated with applications from different domains.
    To evaluate our approach, we present a prototype implementation of FReD and show the benefits of developing with FReD using three case studies of fog computing applications.
\end{abstract}

\newpage
\input{sections/1_introduction.tex}
\input{sections/2_background.tex}
\input{sections/3_relwork.tex}
\input{sections/4_design.tex}
\input{sections/5_evaluation.tex}
\input{sections/6_discussion.tex}
\input{sections/7_conclusion.tex}

% \ack
\section*{Acknowledgement}
Funded by the Deutsche Forschungsgemeinschaft (DFG, German Research Foundation) -- 415899119.

\bibliographystyle{./ACM-Reference-Format}
\bibliography{bibliography.bib}

\end{document}

%% file: sections/1_introduction.tex
\section{Introduction}
\label{sec:introduction}

In fog computing, leveraging the combined resources of the cloud, geo-distributed edge nodes, and intermediary core network infrastructure can provide the necessary Quality of Service (QoS) improvements to facilitate new application domains~\cite{Bonomi2012-if,Zhang2015-cb,Shi2016-pb,Dastjerdi2016-xd,paper_bermbach_fog_vision}.

While this is an exciting new opportunity, developing applications on fog infrastructure is non-trivial, as architectures need to be adapted or completely re-designed for an increasing degree of geo-distribution compared to today's centralized cloud systems~\cite{paper_bermbach_fog_vision}.
One emerging challenge is that of geo-distributed data management:
To benefit from fog computing in a scalable and cost-efficient manner, data replication on the heterogeneous fog infrastructure needs to be controlled carefully:
Specifically, this aims for (near) local data access while meeting data consistency requirements and regulatory demands of applications~\cite{paper_bermbach_fog_vision,Mor2016-hv,Trivedi2020-qh,paper_pfandzelter_functions_streams,paper_pfandzelter_zero2fog}.
Reimplementing systems that meet this demand from scratch for every fog application places a tremendous and unnecessary burden on application developers.

We argue that the missing element in the systems landscape is a replication middleware that distributes data across different sites on the cloud-to-edge continuum and uses the most suitable approach for data storage and query processing on each site.
To this end, we further develop concepts of previous work~\cite{paper_hasenburg_towards_fbase} and present \emph{FReD} (\emph{F}og \emph{Re}plicated \emph{D}ata), a middleware that can serve as the building block for application-configurable fog data replication and distribution.
Unlike existing data management platforms, FReD is tailored for the fog yet application-agnostic, configurable, and extensible\footnote{Our prototype implementation of FReD is available as open-source: \url{https://github.com/OpenFogStack/FReD}}.

To this end, we make the following core contributions in this paper:

\begin{itemize}
    \item We identify the unique requirements for fog data replication systems (\cref{sec:background}), and we discuss where existing approaches fall short (\cref{sec:relwork}).
    \item We present the design of FReD and its abstractions (\cref{sec:design})\footnote{We published the initial ideas behind this system in~\cite{paper_hasenburg_towards_fbase}.}, including our approach to data consistency and our prototype implementation.
    \item We evaluate FReD based on our open-source prototype implementation and analyze to which degree it meets our requirements (\cref{sec:evaluation}).
    \item Finally, we discuss the limitations of this work and present future research directions in fog data management (\cref{sec:discussion}).
\end{itemize}

%% file: sections/2_background.tex
\section{Background \& Requirements}
\label{sec:background}

In this section, we give a brief overview of the field of fog computing and derive requirements for fog data replication.
We also introduce the terminology we use in the rest of this work.
Particularly the term ``fog computing'', as well as competing terms ``multi-access edge computing'', ``mobile edge computing'', ``hierarchical cloud computing'', or ``mist computing'', have been used with varying definitions.
In this paper, we adopt the definition of Bermbach et al.~\cite{paper_bermbach_fog_vision}, where fog computing is the combination of resources in cloud, edge, and core network to provide compute and storage resources across the cloud-edge continuum.
This definition explicitly excludes user equipment and devices from the fog, which we refer to as \emph{clients} accessing the fog infrastructure.

\begin{figure}
    \centering
    \includegraphics[width=\columnwidth]{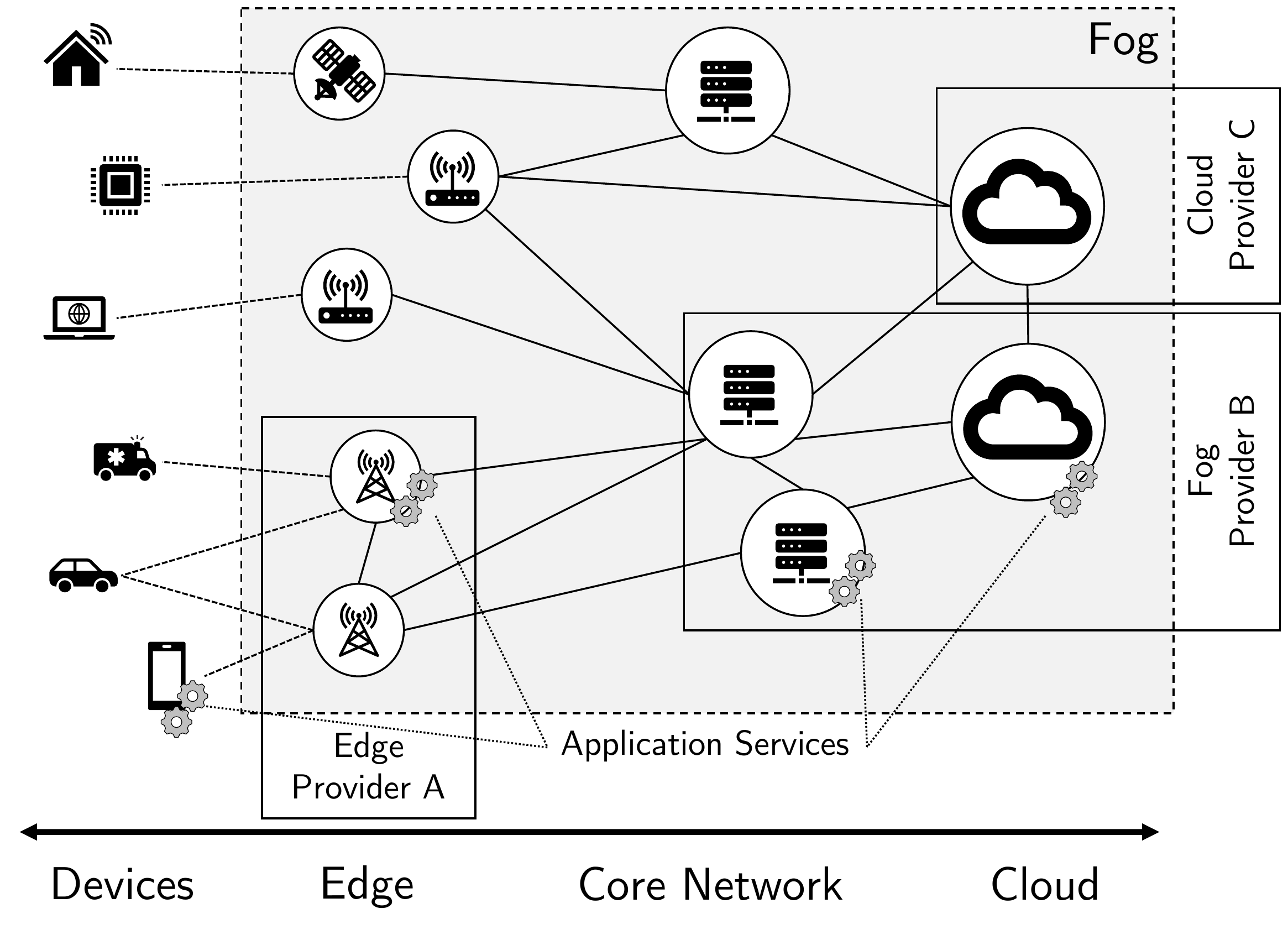}
    \caption{\textbf{Components in Fog Infrastructure~\cite{paper_bermbach_fog_vision}:} Client devices communicate with nodes at the edge of the network. Intermediate nodes within the core network can provide additional services, and cloud datacenters provide compute and storage resources with seemingly infinite scalability. The components may be offered by different providers, and application services are deployed on a subset of these components.}
    \label{fig:foginfrastructure}
\end{figure}

We show an overview of the components of fog computing infrastructure in \cref{fig:foginfrastructure}.
We note that although fog infrastructure is often illustrated as a hierarchical tree with the cloud as its root and edge nodes as leaves, e.g., by Mortazavi et al.~\cite{Mortazavi2017-cu}, this is merely an abstracted overlay topology.
We assume that in practice and with sufficient firewall configuration, servers on the Internet are able to connect and send messages to any other server~\cite{Zhang2015-cb}, possibly with varying QoS~\cite{paper_bermbach_grassroots_peering_edge}.

\subsection{Fog Infrastructure}
\label{sec:background:infrastructure}

\begin{figure}
    \centering
    \includegraphics[width=\columnwidth]{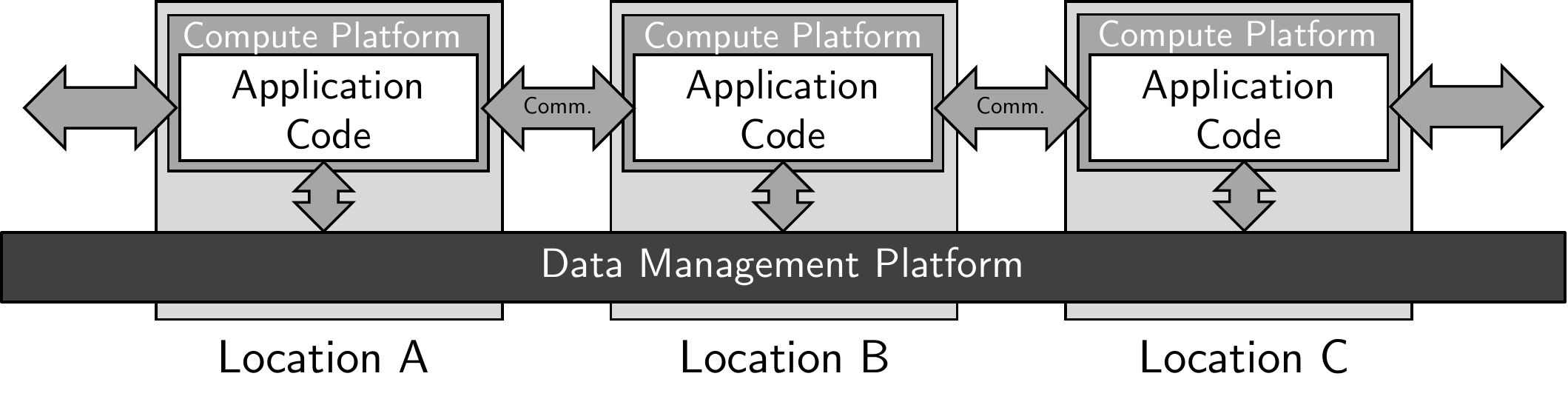}
    \caption{\textbf{Supporting Layers for Fog Applications:} Application code runs across geo-distributed compute locations using a compute platform at each location. Communication channels facilitate message distribution across application instances. The fog data distribution platform replicates data across the geo-distributed locations transparently and can be accessed by instances of the application at each location.}
    \label{fig:foglayers}
\end{figure}

As fog computing is an extension of cloud computing, tenants will expect a cloud-like experience, especially regarding ease-of-use.
Unlike cloud computing, however, fog computing infrastructure is heterogeneous and highly geo-distributed~\cite{Shi2016-zu,paper_bermbach_fog_vision}.
As a result, supporting platform technologies are necessary to enable developers of geo-distributed applications to efficiently take advantage of fog computing.
Specifically, there are three such layers, as shown in \cref{fig:foglayers}: a \emph{compute platform} that runs application code, \emph{communication channels} to communicate across geo-distributed locations, and a \emph{data management middleware} for replication and persistence of application-specific data.
The compute platform could, e.g., be an edge-based FaaS platform such as tinyFaaS~\cite{paper_pfandzelter_tinyfaas} or NanoLambda~\cite{paper_george_nanolambda}, a Kubernetes\footnote{\url{https://kubernetes.io}} cluster or FaaS service in the cloud, or an edge orchestrator such as K3s\footnote{\url{https://k3s.io}} on medium-sized fog nodes.
For communication with other nodes, the application could rely on various \textit{communication channels}, e.g., via a local MQTT broker~\cite{paper_hasenburg_broadcast_groups} or a sidecar proxy.

The fog data management middleware, which we focus on in this paper, must be able to manage both geo-distribution and heterogeneity in the fog.
Heterogeneity is caused by two factors:
First, fog infrastructure comprises infrastructure of different providers, from cloud service providers to telecommunications companies collocating compute resources with their network equipment.
As a result, different nodes in the system can have different hardware architectures, operating systems, or availability guarantees, or the offered platforms and services may be completely distinct~\cite{Bonomi2012-if,Mor2016-hv,Gadepalli2019-bz}.
Applications that manage data in the fog should not be confronted with software stack heterogeneity, e.g., in the form of different database technologies on different nodes
This leads us to our first requirement for a fog data distribution middleware, namely that applications should instead be able to use a \emph{common interface for data access} (\emph{RQ1}).
Second, nodes towards the cloud-end of the fog continuum can be expected to offer more cost-efficient resources and higher scalability and availability.
Cloud data centers and, to a lesser extent, intermediary nodes can benefit more from economies of scale as resources are more centralized than at the edge~\cite{Fox2009-ca}.
A fog data replication middleware needs to keep these costs and performance penalties in mind when replicating and distributing data and should thus \emph{adapt to different node capabilities in the fog} (\emph{RQ2}).

Geographical distribution is a key advantage of fog computing infrastructure, as it enables new kinds of applications where data and services are available close to end-user devices.
It is, however, also a new challenge for fog computing users, as this geo-distribution has to be managed effectively.
While developers can already choose between cloud data center regions in different countries, fog nodes offer finer granularity.
Depending on the infrastructure, this choice may come down to different street corners in the same city~\cite{Rausch2020-cm}.
Our fog data replication middleware should \emph{directly reflect this high level of geo-distribution} (\emph{RQ3}), the opposite of the transparency requirements of traditional distributed systems~\cite{Tanenbaum2016-jp}.

\subsection{Fog Applications}
\label{sec:background:applications}

Fog computing brings increased efficiency for applications and can even enable new application domains that were not attainable solely with cloud infrastructure.
We present three kinds of applications that can benefit from the paradigm of fog computing: delay-sensitive applications~(\cref{sec:background:applications:delay}), bandwidth-intensive applications~(\cref{sec:background:applications:bandwidth}), and privacy-conscious applications~(\cref{sec:background:applications:privacy}).
We note that an application or its components can also fall into multiple or all of these classes, e.g., being both privacy-conscious and delay-sensitive.

\subsubsection{Delay-Sensitive Applications}
\label{sec:background:applications:delay}

A commonly mentioned advantage of fog computing is low latency~\cite{paper_pfandzelter_functions_streams,Mohan2020-cn}.
Processing client data and events on a nearby edge node reduces network latency compared to sending that data to a distant cloud data center, in most cases to a single network hop~\cite{Zhang2015-cb,paper_pfandzelter_functions_streams}.
If the client software requires short reaction times from the services it accesses, cloud computing is not an option:
Even if the cloud can provide an acceptable \emph{average} network delay, percentile latency can be significantly higher given that data need to traverse parts of the wider Internet, subject to, e.g., routing errors~\cite{Mohan2020-cn}.
While processing delay can become more of an issue given the difference in compute capabilities between edge and cloud, this is less of a challenge for simple event processing use cases where the service only needs to quickly react to single events~\cite{Zhang2015-cb,paper_pfandzelter_functions_streams,paper_pfandzelter_tinyfaas}.
To enable these kinds of applications, a fog data management system should \emph{offer low-hop data access} (\emph{RQ4}) where needed.

\subsubsection{Bandwidth-Intensive Applications}
\label{sec:background:applications:bandwidth}

Processing data close to clients on fog infrastructure also provides bandwidth benefits for data-intensive applications.
Sending all data, e.g., measurements of IoT sensors, to the cloud, can be prohibitively expensive or even impossible, and preprocessing in the form of filtering or aggregation on fog infrastructure is a viable solution to this issue~\cite{paper_pfandzelter_zero2fog}.
Similarly, if clients consume large quantities of data, replicating that data to a suitable edge node first and serving this replica to clients can reduce data access costs~\cite{Zhang2015-cb,paper_pfandzelter_functions_streams,paper_pfandzelter_LEO_CDN}.
This requires the replication middleware to \emph{directly integrate with processing services} (\emph{RQ5}) for early aggregation and filtering.

\subsubsection{Privacy-Conscious Applications}
\label{sec:background:applications:privacy}

While fog computing's multi-provider, geo-distributed infrastructure introduces additional technical challenges when it comes to data security and system safety, it does provide opportunities for increased privacy in fog applications~\cite{paper_bermbach_fog_vision,paper_pallas_fog4privacy}.
With governments introducing legislation such as the GDPR, controlling data placement locations becomes a necessary step in managing any application.
Fog computing enables finely-grained, compliant data placement across a geo-distributed infrastructure, such that data can be placed in the clients' regions.
Early data aggregation on fog infrastructure can also increase user privacy as additional steps to anonymize data can be performed on distributed infrastructure that is not controlled by a single cloud provider~\cite{Alrawais2017-xh,grambow_public_2018,Rao2019-xt,paper_pallas_fog4privacy}.
Where needed for compliance and privacy, applications should be able to \emph{directly control data movement in the fog} (\emph{RQ6}).

\subsection{Challenges of Fog Computing Adoption}
\label{sec:background:adoption}

Although the fog computing paradigm is promising, several challenges still lie ahead in broader fog computing adoption.

As of now, pay-as-you-go fog infrastructure is not available in the same capacity as cloud resources.
While some edge compute and storage services are available for integration into applications, we are still missing nodes in direct proximity to clients, e.g., at cellular towers.
This lack of existing infrastructure leaves many assumptions about fog infrastructure to speculation, e.g., the level of geo-distribution or infrastructure heterogeneity that applications and platforms must be able to deal with~\cite{paper_bermbach_fog_vision}.
Fog platforms that are designed now should thus be \emph{agnostic of the fog topology} (\emph{RQ7}) to stay relevant and useful as fog infrastructure is developed and evolves over time.

As fog infrastructure is not built out sufficiently yet, practitioners are also unable to analyze geo-distributed applications for their commercial viability.
The applications that researchers are currently expecting to benefit from fog infrastructure can thus only be hypothetical.
It follows that our fog data replication middleware should also be \emph{application-agnostic} (\emph{RQ8}) as we cannot know now which applications will leverage fog computing in the future.

Building such an application for any heterogeneous, geographically dis\-tri\-but\-ed infrastructure places a huge burden on application developers.
Managing infrastructure from different providers, with different software and hardware stacks, and across hundreds if not thousands of disperse compute nodes is far more difficult than managing only centralized cloud infrastructure.
In many aspects, fog computing is still missing the right abstractions to facilitate efficient development of such applications~\cite{paper_pfandzelter_zero2fog,paper_bermbach_fog_vision}.
Consequently, a fog data replication middleware needs to \emph{manage data consistency and availability} (\emph{RQ9}) for its applications

\subsection{Requirements}
\label{subsec:requirements}

We therefore arrive at the following requirements for a fog data replication middleware:

\paragraph{RQ1: Common Data Access Interface}
Applications require a common interface for data access across infrastructure from different providers, regardless of the available hardware and platforms.

\paragraph{RQ2: Adapt to Infrastructure Heterogeneity}
Infrastructure heterogeneity should be abstracted for the applications as much as possible, while the middleware adapts to the different capabilities on the edge-cloud continuum.

\paragraph{RQ3: Transparent Geo-Distribution}
Geo-distribution of data should be accurately reflected in the middleware and information on data location exposed to applications.

\paragraph{RQ4: Low-Hop Data Access}
The middleware should not stand in the way of building latency-sensitive applications by enabling low-hop data access when and where needed.

\paragraph{RQ5: Integrate Aggregation and Filtering}
Early processing, filtering, and aggregation are required to build efficient data-intensive fog applications, and a fog data replication middleware should make a seamless integration of such processing possible.

\paragraph{RQ6: Control Data Movement}
Data distribution and replication in our middleware should be controllable by the application where required for sensitive information and regulatory compliance.

\paragraph{RQ7: Topology-Agnostic}
As fog computing infrastructure is still being developed and implemented, no assumptions other than heterogeneity and geo-distribution about future fog topology should influence the design of a fog data middleware.

\paragraph{RQ8: Application-Agnostic}
Designing fog data management for a specific type of application limits its use for future applications that might benefit from fog computing, and it is thus important that the middleware provides a use-case-agnostic interface.

\paragraph{RQ9: Manage Data Consistency and Availability}
Managing data replication and distribution alone is not sufficient, as data consistency and availability over the distributed fog infrastructure is a large burden for applications.

%% file: sections/3_relwork.tex
\section{Related Work}
\label{sec:relwork}

In this section, we give an overview of existing work in distributed data management and discuss these approaches regarding their suitability for fog data management in light of our requirements.
We start with existing solutions from the fields of cloud and grid computing, where managing distributed data has long been a research subject.
These approaches can be classified based on their replica placement strategies: \emph{global mapping} (\cref{sec:relwork:mapping}), \emph{hashing} (\cref{sec:relwork:hashing}), and \emph{scattering} (\cref{sec:relwork:scattering})~\cite{MacCormick2009-rc}.
For each of the categories, we discuss examples.
Afterwards, we discuss existing fog-native solutions that were specifically designed for geo-distributed, heterogeneous infrastructure and fog applications (\cref{sec:relwork:fognative}).
We show an overview of the state-of-the-art in distributed data management that we discuss in this section in \cref{tab:relwork}.

\input{tables/relwork.tex}

\subsection{Global Mapping}
\label{sec:relwork:mapping}

In a global mapping approach, replica placement is controlled by a single, centralized component.
In deployments, this logical single entity is typically replicated to increase availability.
While this supports arbitrary complex and intelligent replica placement decisions, such systems come with natural scalability and availability challenges, e.g., introducing a single point of failure, or leading to bottlenecks as all control flow needs to pass through a centralized component.
A well-known example of a storage system using global mapping is
the \emph{Google File System} (GFS)~\cite{Ghemawat2003-gg} which also served as a blueprint for the development of HDFS in the Hadoop ecosystem~\cite{hadoopfilesystem}.
In GFS, a single master server handles replica placement and selection across the entire (not geo-distributed) GFS cluster, while shadow master servers help to improve availability.
As another example, Nebula~\cite{Ryden2014-ow}, a grid-inspired distributed edge store, centrally controls data placement in a \emph{DataStore Master} that handles all aspects of data placement and replica selection.
In a large-scale geo-distributed deployment, however -- for both GFS and Nebula -- directing all requests first to such a centralized master server is prohibitive both in terms of performance, scalability, and availability.
Similarly, keeping a geo-replicated master synchronized just shifts the replication and tradeoff problems to the cluster management level.

\subsection{Hashing}
\label{sec:relwork:hashing}

In hashing-based replica placement, a set of machines to store a data item is deterministically identified by a hash value, usually of the data item's key.
As this fully decentralized approach for replica placement control and selection scales well, it was chosen in a large variety of systems.
Typically, it comes in one of two flavors:
Systems such as Chord~\cite{Stoica2001-by}, PAST~\cite{Druschel2001-sc}, Pastry~\cite{Rowstron2001-qi}, Oceanstore~\cite{Kubiatowicz2000-he}, and Kademlia~\cite{Maymounkov2002-ra} use multi-hop routing to locate data items
More recent systems, e.g., Cassandra~\cite{Lakshman2010-zm}, Voldemort~\cite{sumbaly2012serving}, and other Dynamo-inspired~\cite{DeCandia2007-ib} data stores organize their machines on a ring structure controlling the assignment of hash ranges to machines.
Overall, while hashing-based systems scale well and are relatively easy to implement, they often do not cope well with high node churn rates.
Furthermore, they are not a particularly good fit for geo-distributed environments in general but specifically not for fog deployments as the full determinism of the static hash function makes it hard to consider underlying network topologies in replica placement.
Dynamic replica placement, i.e., placing data close to actual access locations based on current demand, is also not possible due to the static nature of the hash function.

Dynamo, Cassandra, and MetaStorage~\cite{paper_bermbach_metastorage} further extend hashing with chaining, where additional replicas of a data item on nodes adjacent to the primary replica.
Dynamo places additional replicas on the next \emph{N-1} nodes after the primary replica, as determined by consistent hashing, on its ring structure.
In case of node failure, this is relaxed to the next N-1 \emph{healthy} nodes, a mechanism called hinted handoffs.
While chaining could consider network topology in a fog environment, its static nature cannot support dynamic data movement.

\subsection{Scattering}
\label{sec:relwork:scattering}

Scattering is similar to hashing in that a deterministic, static functions assigns data items to replicas.
CRUSH~\cite{Weil2006-pp}, as used, e.g., in Ceph~\cite{Weil2006-qs}, uses a pseudorandom distribution of replicas across nodes based on a hierarchical description of the target cluster.
In contrast to hashing, this enables administrators to dynamically add and remove nodes without shifting hash ranges across the cluster.
Nevertheless, the static nature of the deterministic pseudorandom function makes such systems poorly equipped for mobile clients that require replica movement.

\subsection{Fog-Native}
\label{sec:relwork:fognative}

We have seen that existing grid and cloud approaches for data management are a poor fit for the unique characteristics of fog environments.
Hence, numerous fog-native data management solutions have been proposed~\cite{10.1002/9781119525080}.
Hourglass~\cite{Shneidman2004-km} introduces \emph{circuits}, an abstraction for data flow from sensor networks to data consumers and in-network services.
This approach, however, limits its applicability for modern fog applications as circuits are unidirectional.

Lin et al.~\cite{Lin2007-if} propose full database replication on the edge to enable low-latency data access with strong consistency guarantees.
Such an approach is unlikely to work with high degrees of geo-distribution across hundreds or thousands of nodes and cannot consider infrastructure heterogeneity.

With the \emph{Global Data Plane} (GDP)~\cite{Zhang2015-cb}, Zhang et al.~present a data management system for the IoT around single-writer, append-only logs that can be migrated across heterogeneous compute platforms and connected to build applications.
This approach is particularly well-suited for the IoT that produces large amounts of data, but cannot serve as an application-agnostic platform as it does not support mutable data.

Confais et al.~\cite{Confais2017-do,Confais2017-bc,Confais2020-um} analyze the fog-readiness of several storage systems and then extend the most suitable one, the interplanetary filesystem (IPFS), with a scale-out NAS in an effort to make it more fog-ready.
This exposes data locality to some extent and gives clients local read and write access.
Nevertheless, as IPFS only supports immutable files, this system is not application-agnostic.

\emph{Cloudpath}~\cite{Mortazavi2017-cu,mortazavi2020feather,Mortazavi2020-ah} is a framework for a hierarchical fog computing infrastructure that includes a data management system, \emph{PathStore}.
In PathStore, storage nodes are arranged in a tree topology, with each node running a Cassandra cluster proportioned for its capabilities.
The hierarchical topology allows PathStore to implement a replication strategy where parent nodes replicate all data from their children at row granularity.
On queries, nodes first pull relevant data from their parents, with a query cache preventing pulling data for each access.
The downside of PathStore is that it relies on a tree topology, which is not necessarily representative of fog infrastructure~\cite{paper_bermbach_grassroots_peering_edge}.
Such a tree can, e.g., only have a single root node, whereas several  cloud data centers might be part of the fog environment as equals.
Furthermore, a single network partition leads to an entire subtree losing access to PathStore.

Naas et al.~\cite{Naas2017-ln,Naas2019-qv} present \emph{iFogStor}, which is not a data management system but an approach to scheduling data replicas across distributed, heterogeneous fog infrastructure to minimize data access latency for clients.
The authors propose a heuristic approach based on geographical zoning that considers network characteristics and node capabilities and show that such an approach is feasible at runtime and leads to QoS improvements.
Such an approach, however, is application-specific, in this case to the IoT, and requires a central mapping server to make placement decisions.
Given both privacy concerns and availability requirements, it is thus unlikely to be feasible in practice.

FogStore~\cite{Mayer2017-ti,Gupta2018-ha,Gupta2018-rl} is a geo-distributed key-value store for the fog that provides low latency data access with differential consistency guarantees.
FogStore introduces the notion of a \emph{Context of Interest} of data items that is configurable by applications.
Strong consistency is only guaranteed if the client accessing a data point is located within this context of interest, which foregoes full, strongly consistent replication across the entire system.
This approach fits well with the notion of IoT devices producing data with correlated information about their geographical position that provides important context to data (cf.~Hasenburg et al.~\cite{paper_hasenburg_geocontext}).
Mobile data consumers, on the other hand, cannot control movement of their data, which limits the applicability of FogStore for a broad range of fog applications.

Plebani et al.~\cite{paper_ditas} introduce the idea of \emph{Virtual Data Containers} that group coherent data items.
These containers can then be moved around depending on where data is needed, providing a flexible interface for controlled data movement through the fog.
This idea has not been developed into a data management system yet, and many questions are left unanswered, e.g., how data containers could be replicated to constrained edge nodes.

Psaras et al.~\cite{Psaras2018-ub} propose mobile data repositories at the edge that are provided per-user and allow clients to upload and retrieve data.
These mobile data repositories are supposed to move with the client and to be augmented with eventually consistent replication to the cloud and intermediary processing services.
This approach is tailored for data that is bound to a specific client and should be replicated from the edge to a central cloud, yet it does not support replication along the edge of the network which would give multiple clients low-latency access to the same data set.

Trivedi et al.~\cite{Trivedi2020-qh} outline \emph{Griffin}, a distributed storage service for the fog.
While the authors provide only a sketch for their design and use it to propose further research questions, the proposed architecture includes a central replica placement and data creation service.
Although clients can access data directly at edge nodes after coordinating with this central service, it does raise availability and privacy concerns.
Additionally, it is unclear how Griffin deals with client mobility, especially in light of multiple clients accessing the same data.

%% file: tables/relwork.tex
\begin{table}[!t]
    \renewcommand{\arraystretch}{1.3}
    \centering
    \caption{We give an overview of state-of-the-art fog data management approaches as well as the more traditional examples discussed in this section and show to which degree they meet our requirements for a fog data management platform. For comparison, we also show which requirements may be satisfied by building a custom application-specific data management solution. $\triangle$ indicates partial support.}
    \label{tab:relwork}

    \resizebox{\textwidth}{!}{
        \begin{tabular}{cc|c|c|c|c|c|c|c|c|c|}
            \cline{3-11}
                                                                                                                &                                                                                                                                         & \multicolumn{9}{c|}{Requirements}                                                                                                                         \\ \hline
            \multicolumn{1}{|c}{Type}                                                                           & \multicolumn{1}{|c|}{Approach}                                                                                                          & \textbf{RQ1}                      & \textbf{RQ2} & \textbf{RQ3} & \textbf{RQ4} & \textbf{RQ5} & \textbf{RQ6} & \textbf{RQ7} & \textbf{RQ8} & \textbf{RQ9} \\ \hline
            \multicolumn{1}{|c}{\multirow{2}{*}{\begin{tabular}[c]{@{}c@{}}Global\\[-1ex]Mapping\end{tabular}}} & \multicolumn{1}{|c|}{\begin{tabular}[c]{@{}c@{}}\emph{Google File}\\[-1ex]\emph{System}~\cite{Ghemawat2003-gg}\end{tabular}}            & $\checkmark$                      &              &              &              & $\triangle$  &              &              & $\checkmark$ & $\checkmark$ \\ \cline{2-11}
            \multicolumn{1}{|c}{}                                                                               & \multicolumn{1}{|c|}{\emph{Nebula}~\cite{Ryden2014-ow}}                                                                                 & $\checkmark$                      & $\checkmark$ & $\checkmark$ & $\triangle$  & $\checkmark$ &              & $\checkmark$ & $\checkmark$ & $\triangle$  \\ \hline
            \multicolumn{1}{|c}{\multirow{7}{*}{Hashing}}                                                       & \multicolumn{1}{|c|}{\emph{Chord}~\cite{Stoica2001-by}}                                                                                 & $\checkmark$                      &              &              &              &              &              & $\checkmark$ & $\checkmark$ & $\checkmark$ \\ \cline{2-11}
            \multicolumn{1}{|c}{}                                                                               & \multicolumn{1}{|c|}{\emph{Past}~\cite{Druschel2001-sc}, \emph{Pastry}~\cite{Rowstron2001-qi}}                                          & $\checkmark$                      &              &              & $\triangle$  &              &              & $\checkmark$ & $\checkmark$ &              \\ \cline{2-11}
            \multicolumn{1}{|c}{}                                                                               & \multicolumn{1}{|c|}{\emph{Kademlia}~\cite{Maymounkov2002-ra}}                                                                          & $\checkmark$                      &              &              &              &              &              & $\checkmark$ & $\checkmark$ & $\checkmark$ \\ \cline{2-11}
            \multicolumn{1}{|c}{}                                                                               & \multicolumn{1}{|c|}{\emph{Oceanstore}~\cite{Kubiatowicz2000-he}}                                                                       & $\checkmark$                      &              & $\triangle$  & $\triangle$  & $\triangle$  & $\triangle$  & $\checkmark$ & $\checkmark$ & $\checkmark$ \\ \cline{2-11}
            \multicolumn{1}{|c}{}                                                                               & \multicolumn{1}{|c|}{\emph{Cassandra}~\cite{Lakshman2010-zm}}                                                                           & $\checkmark$                      &              &              &              &              &              &              & $\checkmark$ & $\checkmark$ \\ \cline{2-11}
            \multicolumn{1}{|c}{}                                                                               & \multicolumn{1}{|c|}{\emph{Dynamo}~\cite{DeCandia2007-ib}}                                                                              & $\checkmark$                      &              &              &              &              &              &              & $\checkmark$ & $\checkmark$ \\ \cline{2-11}
            \multicolumn{1}{|c}{}                                                                               & \multicolumn{1}{|c|}{\emph{MetaStorage}~\cite{paper_bermbach_metastorage}}                                                              & $\checkmark$                      &              & $\triangle$  & $\checkmark$ &              & $\triangle$  &              & $\checkmark$ & $\checkmark$ \\ \hline
            \multicolumn{1}{|c}{\multirow{1}{*}{Scattering}}                                                    & \multicolumn{1}{|c|}{\begin{tabular}[c]{@{}c@{}}\emph{CRUSH}~\cite{Weil2006-pp},\\[-1ex]\emph{Ceph}~\cite{Weil2006-qs}\end{tabular}}    & $\checkmark$                      & $\triangle$  &              &              &              &              &              & $\checkmark$ & $\checkmark$ \\ \hline
            \multicolumn{1}{|c}{\multirow{13}{*}{Fog-Native}}                                                   & \multicolumn{1}{|c|}{\emph{Hourglass}~\cite{Shneidman2004-km}}                                                                          & $\checkmark$                      & $\checkmark$ & $\checkmark$ &              & $\checkmark$ & $\triangle$  & $\triangle$  &              & $\triangle$  \\ \cline{2-11}
            \multicolumn{1}{|c}{}                                                                               & \multicolumn{1}{|c|}{Lin et al.~\cite{Lin2007-if}}                                                                                      & $\checkmark$                      &              &              & $\checkmark$ &              &              & $\checkmark$ & $\checkmark$ & $\checkmark$ \\ \cline{2-11}
            \multicolumn{1}{|c}{}                                                                               & \multicolumn{1}{|c|}{\begin{tabular}[c]{@{}c@{}}\emph{Global Data}\\[-1ex]\emph{Plane}~\cite{Zhang2015-cb}\end{tabular}}                & $\triangle$                       & $\checkmark$ & $\checkmark$ & $\triangle$  &              &              & $\checkmark$ &              & $\checkmark$ \\ \cline{2-11}
            \multicolumn{1}{|c}{}                                                                               & \multicolumn{1}{|c|}{\begin{tabular}[c]{@{}c@{}}fog-ready\\[-1ex]IPFS~\cite{Confais2017-do,Confais2017-bc,Confais2020-um}\end{tabular}} & $\checkmark$                      & $\triangle$  & $\checkmark$ & $\triangle$  &              & $\triangle$  & $\triangle$  & $\triangle$  & $\triangle$  \\ \cline{2-11}
            \multicolumn{1}{|c}{}                                                                               & \multicolumn{1}{|c|}{\emph{PathStore}~\cite{Mortazavi2017-cu,mortazavi2020feather,Mortazavi2020-ah}}                                    & $\checkmark$                      & $\checkmark$ & $\checkmark$ & $\triangle$  & $\checkmark$ &              &              & $\checkmark$ & $\triangle$  \\ \cline{2-11}
            \multicolumn{1}{|c}{}                                                                               & \multicolumn{1}{|c|}{Naas et al.~\cite{Naas2017-ln,Naas2019-qv}}                                                                        &                                   & $\checkmark$ & $\triangle$  & $\checkmark$ &              &              &              & $\triangle$  & $\checkmark$ \\ \cline{2-11}
            \multicolumn{1}{|c}{}                                                                               & \multicolumn{1}{|c|}{\emph{FogStore}~\cite{Mayer2017-ti,Gupta2018-rl,Gupta2018-ha}}                                                     & $\checkmark$                      & $\triangle$  & $\checkmark$ & $\checkmark$ &              & $\triangle$  & $\checkmark$ & $\triangle$  & $\checkmark$ \\ \cline{2-11}
            \multicolumn{1}{|c}{}                                                                               & \multicolumn{1}{|c|}{\begin{tabular}[c]{@{}c@{}}Virtual Data\\[-1ex]Containers~\cite{paper_ditas}\end{tabular}}                         & $\checkmark$                      &              & $\checkmark$ & $\triangle$  & $\triangle$  & $\checkmark$ & $\checkmark$ & $\checkmark$ &              \\ \cline{2-11}
            \multicolumn{1}{|c}{}                                                                               & \multicolumn{1}{|c|}{Psaras et al.~\cite{Psaras2018-ub}}                                                                                & $\checkmark$                      & $\triangle$  & $\checkmark$ & $\checkmark$ & $\checkmark$ & $\triangle$  & $\triangle$  & $\checkmark$ &              \\ \cline{2-11}
            \multicolumn{1}{|c}{}                                                                               & \multicolumn{1}{|c|}{\emph{Griffin}~\cite{Trivedi2020-qh}}                                                                              & $\triangle$                       & $\checkmark$ & $\checkmark$ & $\checkmark$ &              &              & $\checkmark$ & $\checkmark$ & $\triangle$  \\ \hline
            \multicolumn{1}{|c}{}                                                                               & \multicolumn{1}{|c|}{\begin{tabular}[c]{@{}c@{}}Custom\\[-1ex]Data Management\end{tabular}}                                             &                                   & $\checkmark$ & $\checkmark$ & $\checkmark$ &              & $\triangle$  & $\checkmark$ & $\triangle$  &              \\ \hline
        \end{tabular}
    }
\end{table}

%% file: sections/4_design.tex
\section{Design \& Implementation}
\label{sec:design}

We have shown that managing data replication and distribution in the fog is a non-trivial problem that has not been sufficiently solved yet
To address this challenge, we present FReD, describing key concepts (\cref{sec:design:concepts}), replication and data distribution (\cref{sec:design:replication}), consistency guarantees (\cref{sec:design:consistency}), and our proof-of-concept implementation (\cref{sec:design:prototype}).

\subsection{FReD Concepts}
\label{sec:design:concepts}

\begin{figure}
    \centering
    \includegraphics[width=\textwidth]{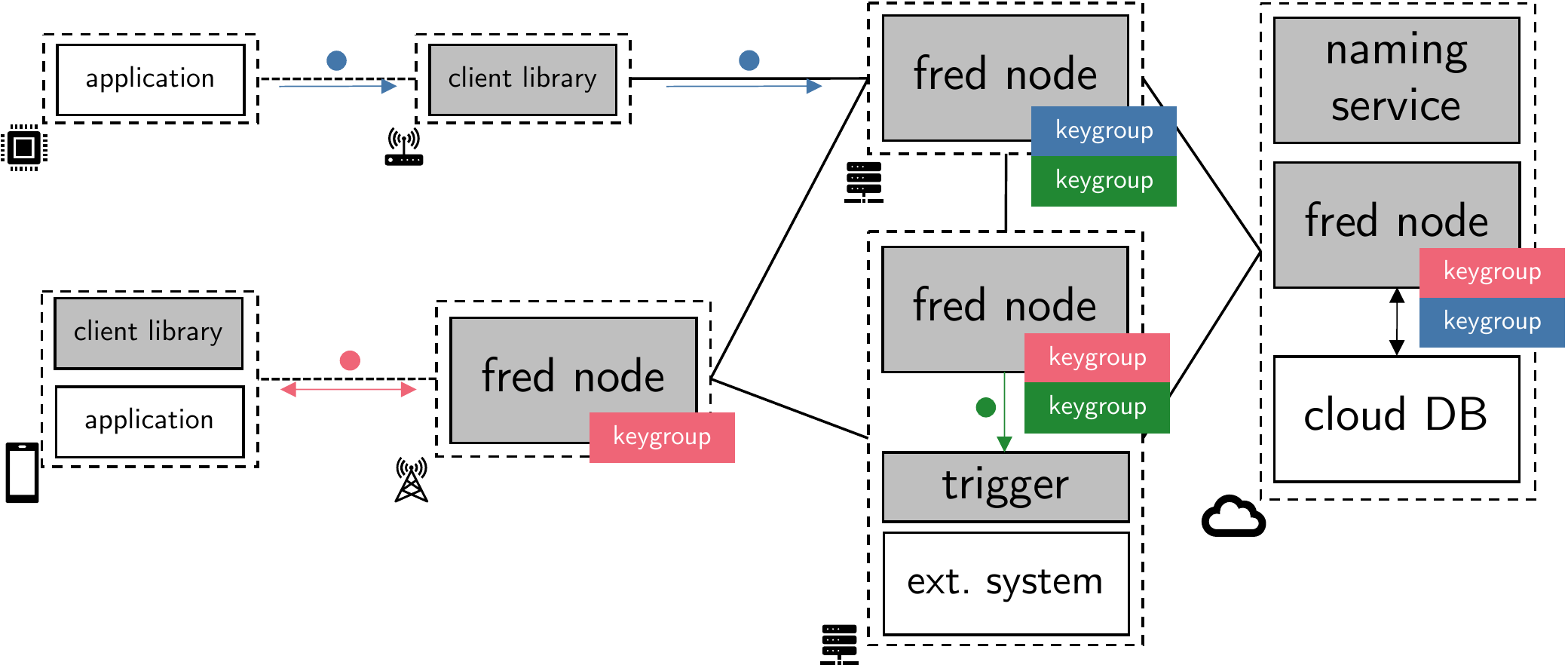}
    \caption{\textbf{Example FReD deployment on distributed fog infrastructure:} A mobile client reads and writes ``\textbf{\textred{red}}'' data to a local fog node through the local client library, an IoT sensor writes ``\textbf{\textblue{blue}}'' data through the library located on a gateway, and an external system is integrated at a fog node through a trigger to process ``\textbf{\textgreen{green}}'' data.}
    \label{fig:fred}
\end{figure}

FReD has several key concepts:
A \emph{node} is a set of machines running FReD at a specific location in the fog.
\emph{Clients} are applications that communicate with nodes using a lightweight local client library.
The \emph{naming service} provides global topology information to nodes.
\emph{Keygroups} are sets of logically coherent key-value pairs which are replicated together and share the same access control properties.
Finally, \emph{triggers} are mechanisms to integrate external systems in an event-driven way.
We show an overview of an example FReD deployment using these concepts in \cref{fig:fred}.

\paragraph*{Keygroups}
Keygroups are FReD's data set objects.
A keygroup groups logically coherent data, e.g., data of a single application user.
Keygroups can be mutable sets of key-value pairs or append-only logs where each entry receives a unique key automatically.
All replication and distribution policies are applied at keygroup granularity, which helps applications manage coherent data as single objects rather than each data item individually.

\begin{figure}
    \centering
    \begin{subfigure}{\textwidth}
        \centering
        \includegraphics[scale=0.6]{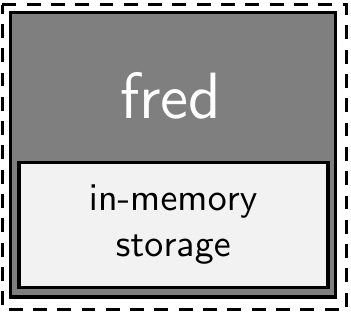}
        \caption{On constrained edge nodes, only in-memory data storage might be possible.}
        \label{fig:fred-deployment-edge}
    \end{subfigure}
    \vfill
    \begin{subfigure}{\textwidth}
        \centering
        \includegraphics[scale=0.6]{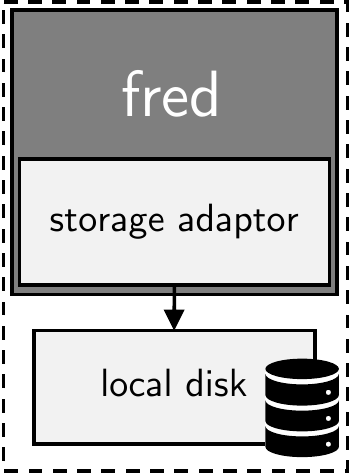}
        \caption{Intermediate nodes with one physical machine can persist data locally on disk.}
        \label{fig:fred-deployment-intermediate}
    \end{subfigure}
    \vfill
    \begin{subfigure}{\textwidth}
        \centering
        \includegraphics[scale=0.6]{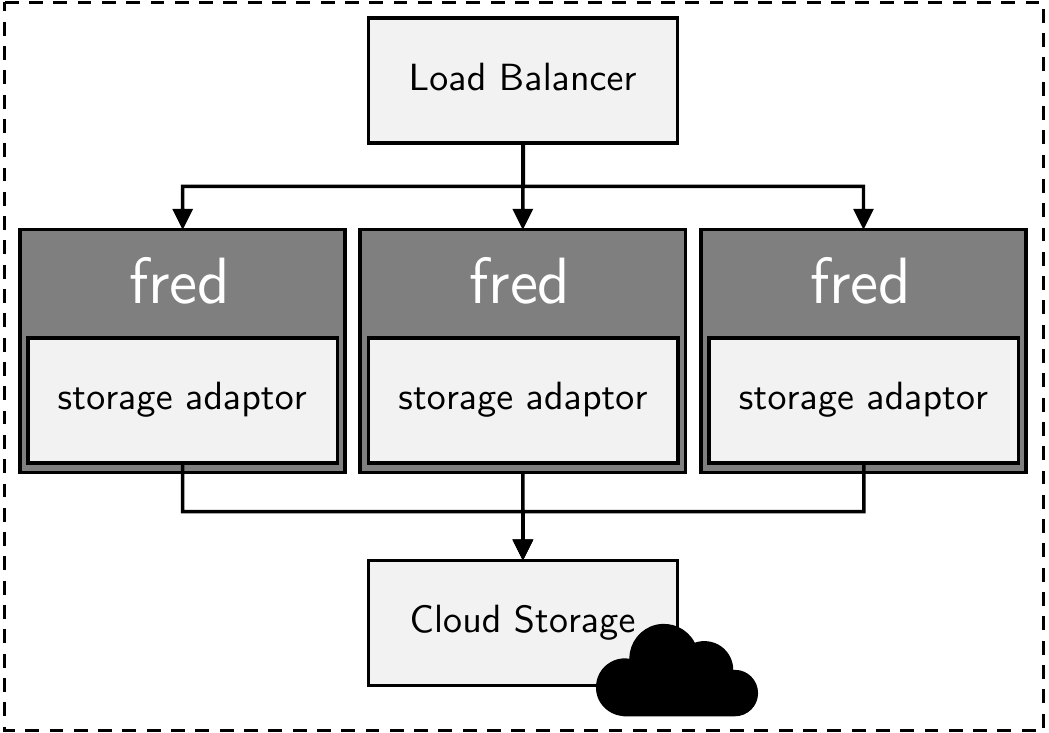}
        \caption{In the cloud, a FReD cluster might share cloud storage for availability and cost.}
        \label{fig:fred-deployment-cloud}
    \end{subfigure}

    \caption{\textbf{Possible FReD Node Deployment Options:} Depending on infrastructure size and capabilities at a particular site, FReD nodes can use different deployment styles. FReD provides a storage abstraction that allows using different storage backends depending on the node type.}
    \label{fig:fred-node-deployments}
\end{figure}

\paragraph*{Nodes}
Nodes in FReD are data storage locations that abstract underlying infrastructure.
A node may comprise a single system-on-a-chip at the edge or a group of cloud virtual machines, as shown in \cref{fig:fred-node-deployments}.
FReD makes no assumptions about the underlying infrastructure except that all nodes can communicate with each other, whether directly or through a load balancer, proxy server, or gateway.
Furthermore, where available, FReD can leverage existing data storage infrastructure such as a cloud database.
As such, all nodes in FReD are considered equal.
Nodes are the replication targets in FReD, which means that a keygroup can be replicated across \emph{N} different nodes.
Nodes not only store data but are also the points of access for clients to perform operations on keygroup data or to change replication and distribution policies.

\paragraph*{Naming Service}
In addition to replicating data, nodes in FReD will infrequently need to exchange configuration data.
This includes a global view of available nodes to aid clients in updating replication policies, share keygroup specific metadata such as access policies, or coordination between physical machines within a node.
Such configuration data can be exchanged in a peer-to-peer manner, e.g., a node informs other members of a keygroup about a new replication target.
In case of conflicts, e.g., concurrent updates, FReD includes a shared naming service that acts as the single source of truth for configuration management.
This naming service does not have any information on data within a keygroup and instead only knows which nodes and keygroups exist and which nodes are replication targets for which keygroups.
An implementation of the naming service can leverage existing centralized or decentralized approaches for fog configuration management~\cite{Jeffery2021-hd,paper_pfandzelter_coordination}.

\paragraph*{Clients}
A FReD client can be an IoT device, mobile handset, web application, or any other software that needs to read or write data in the fog.
Clients can access data at FReD nodes through a local client-side library that provides additional features such as finding optimal nearest nodes, updating replication policies with movement, or providing client-centric consistency guarantees.
For constrained clients such as IoT sensors, such a client library could be a dedicated software component located on a local gateway, while it might be implemented as a software library for web applications.

\paragraph*{Triggers}
Although any system can read data from it, we extend FReD with interfaces to integrate existing systems, which we call triggers.
Triggers are open interfaces that any system can implement, either directly or through a shim service.
They can be added per keygroup to replica nodes and will be called on data updates, e.g., the deletion of a data item from a keygroup will be sent to any external system the developer configures.
This allows building complex data processing pipelines, as the external system's software can of course also act as a further FReD client.

\subsection{Replication \& Distribution}
\label{sec:design:replication}

Replication in FReD is based on a dynamic mapping of keygroups to nodes that is configurable through an API.
We refer to a node that is a replication target for a particular keygroup as a \emph{replica node} for this keygroup.
All replica nodes are informed of all data updates within the keygroup with an optimistic replication approach and can be used to access the data by clients through a uniform CRUD interface.
When a new replica node is added to a keygroup, it will pull existing data from other replica nodes.
As FReD does not distinguish between edge, intermediary, or cloud nodes, keygroups can be configured to be replicated across any non-empty subset of available nodes.
This has the added benefit that replication effort is limited by the number of replica nodes.
It is thus also possible to store privacy-sensitive data only on the edge without any involvement of the cloud, or even to use FReD for data replication between cloud data centers.

Of course, there are differences in available storage and compute capabilities on the edge-cloud continuum.
To account for infrastructure heterogeneity, keygroups can be partially replicated with data expiry.
Constrained replica nodes, e.g., at the edge, can thus be configured to keep only recent data locally available.

The basic abstractions in FReD can be used to build arbitrary data distribution policies.
With replica nodes, applications can define which fog locations should receive their data.
By using data expiry, nodes can also be configured as data entry points for clients that forward their data to other nodes or to triggers.
These triggers can be used for aggregation and filtering, e.g., by using them with edge FaaS functions that push data back into other keygroups.
As keygroup replication can be configured through an API, it can also be dynamically adapted by the application for client mobility:
A client can proactively instruct FReD to replica data to a node that it will connect to in the future.

\subsection{Versioning \& Consistency}
\label{sec:design:consistency}

In FReD, we pay special attention to data consistency, which is a non-trivial problem in distributed systems, and even more so in light of the geo-distributed fog infrastructure.
By default, data in FReD is replicated in an optimistic, eventually consistent manner.
As the PACELC theorem~\cite{Abadi2012-us} states, data replication forces us to decide between data availability and consistency in case of network partitions and between latency and consistency otherwise.
In both cases, we decide on relaxing consistency guarantees to ensure availability and low latency for the applications that require it.
In a fog environment, the heterogeneous infrastructure and communication over the Internet makes network partitions and node failures likely.\footnote{We do note, however, that these partitions and node failures are only relevant for a keygroup when they affect that particular keygroup's replica nodes.}
In addition, we note that data consistency in FReD also depends on the consistency guarantees of the underlying storage system at each node, e.g., if a cloud storage system with eventual consistency is used in a cloud FReD node, FReD can only guarantee eventual consistency for data that is replicated there.

\begin{figure}
    \centering
    \includegraphics[width=\textwidth]{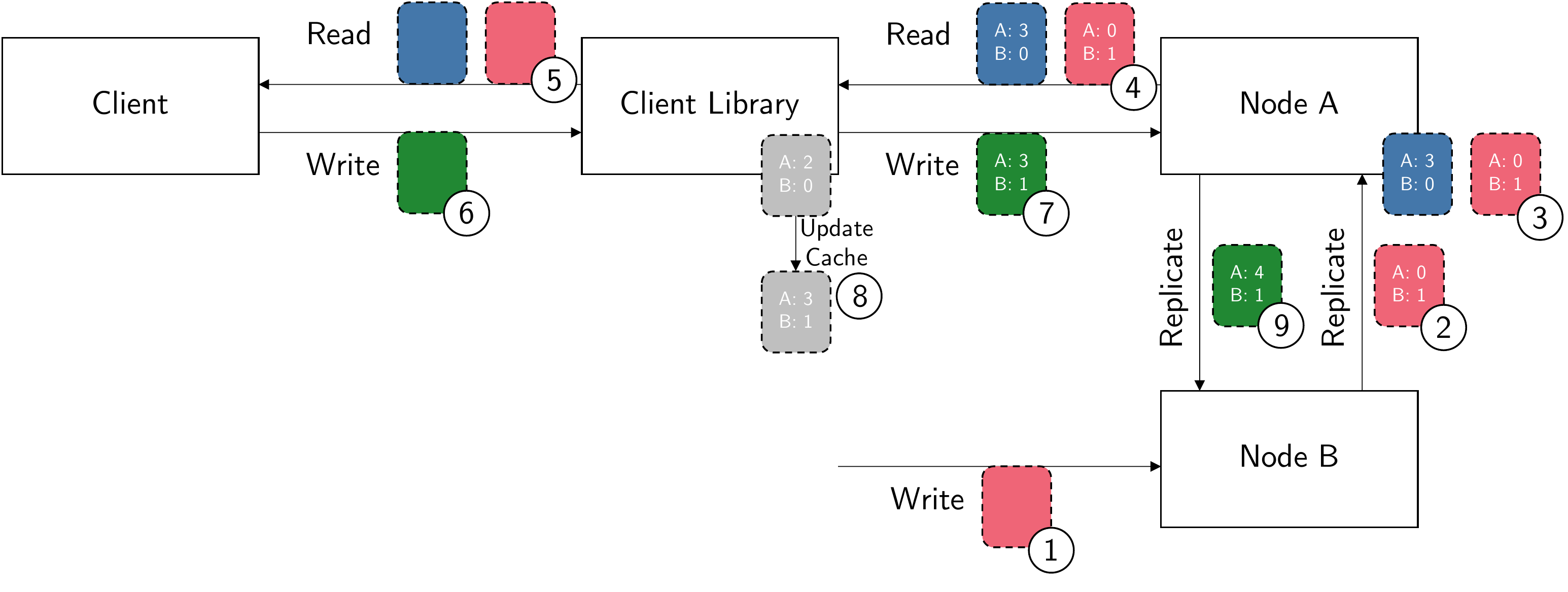}
    \caption{
        \textbf{Example Capturing of a Causal Data Update Relationship with Version Vectors:}
        Version vectors are used internally in FReD to capture causal consistency.
        With a client-side library, FReD can provide client-centric consistency guarantees without exposing versioning to the client.
        Here, a write request for a data item is issued to Node B (\protect\circled{1}), which assigns this value (\textbf{\textred{red}}) the version \textsf{A: 0, B: 1} and replicates it to Node A (\protect\circled{2}).
        Node A has an existing value (\textbf{\textblue{blue}}) for this item with version \textsf{A: 3, B: 0} and keeps both versions, as they are concurrent (\protect\circled{3}).
        The client issues a read request for this data item through the client library, which receives both concurrent values and their versions from Node A (\protect\circled{4}).
        The client library does not expose internal versioning information to the client, instead only sending both concurrent values (\textbf{\textblue{blue}} and \textbf{\textred{red}}, \protect\circled{5}).
        The client can resolve this conflict depending on application logic.
        In this case, the client merges the existing values into a new value (\textbf{\textgreen{green}}) and issues a write request through the client library (\protect\circled{6}).
        The client library then updates this item's version to \textsf{A: 3, B: 1} and writes it to Node A (\protect\circled{7}).
        By updating the local cache with this new version (\protect\circled{8}), the client library knows which item version was last seen to prevent reading old versions on future requests.
        Node A can then also replicate the merged value to Node B, where it supersedes existing values (\protect\circled{9}).
    }
    \label{fig:versioning}
\end{figure}

Nevertheless, we can still achieve a higher level of consistency for applications that require it through client-centric consistency guarantees using our client library.
By supporting version vectors for keys in keygroups, FReD attaches versioning information to data items.
To a client, the client library offers a simple CRUD interface to access data items, yet with a configurable, per keygroup client-centric consistency level.
By tracking read and written versions of data items, the client library can ensure that the required consistency level is achieved, despite the fact that data and their version information is only replicated in an eventually consistent manner by FReD.
We show an overview of this in \cref{fig:versioning}.

For every \emph{Update} operation of a key in a keygroup on a replica node, clients can suggest a set of previously seen version vectors that should be superseded by this operation.
Unless the node locally stores a version that is greater than any of the given versions, a new version will be created for that update by incrementing a local counter~\cite{Tanenbaum2016-jp}.
\emph{Delete} operations follow a similar technique where the value is a tombstone.
Updated values are then replicated across all keygroup replicas with their version vector.
Internally, when a node receives a new value by a peer, it either stores it as a concurrent value or overwrites the existing value, depending on whether the value's version is greater or concurrent to those currently stored.
Any \emph{Read} or \emph{Scan} operation returns a set of values for a particular key in a keygroup.
Each of these values is accompanied by its version vector.

Our client library understands the version vectors used by FReD and the causal ordering they impart, and can apply this to provide client centric consistency guarantees to clients.
For this, the client library keeps a cache of seen versions for all data items that the client accesses.
When a client reads a value from the client library, it is requested from the currently selected FReD node.
If the read yields values with versions greater or equal than those currently in store, they are then returned to the client.
Otherwise, the client library treats outdated values as read failures.
If the read returns multiple values for a key, those values will be given to the client as concurrent values.
This approach guarantees \emph{Monotonic Read Consistency} (MRC), i.e., after reading a version $n$, the client will never again read a version $<n$~\cite{Tanenbaum2016-jp,Vogels2009-oo}.

When a client updates a key, the client library adds the locally cached versions for that key to the update request as it is passed to FReD.
As such, only seen values are overwritten as part of the update.
If multiple concurrent versions are cached, this is treated as conflict resolution, i.e., the new value supersedes all previously seen values.
The resulting version is then also written into the cache, leading to \emph{Read Your Writes Consistency} (RYWC), where the client will never read a version $<n$ after having written version $n$~\cite{Tanenbaum2016-jp,Vogels2009-oo}.
As a replica node rejects updates that carry outdated version vectors, we also get \emph{Monotonic Writes Consistency} (MWC), where two consecutive writes by a client will not lead to a situation where the first overwrites the second write at the same replica~\cite{Tanenbaum2016-jp,Vogels2009-oo}.
In fact, as the client library is part of the storage system from a client perspective, i.e., any requests are made to the same library instance, we can guarantee that writes are serialized in the order that they are made by the client.
This perspective also leads to \emph{Write Follows Read Consistency}, i.e., a write operation following a read of version $n$ only executes on replicas with versions $\ge n$.

In both of these cases, the client itself does not need to deal with any version vectors, as this is all handled by the client library.
Instead, it can simply perform read and write operations as required, but may occasionally have to merge or otherwise deal with concurrent values, which is use-case specific and thus left to client implementation.

Finally, applications may want to bypass the client library and communicate about data versions externally, e.g., through direct communication among two clients.
This leads to external causality that cannot be captured within FReD.
The client library therefore also offers a \emph{Notify} API that lets clients directly update seen versions in the version cache of the client library.

\subsection{Integration into the Fog Application Stack}

\begin{figure}
    \centering
    \includegraphics[width=\textwidth]{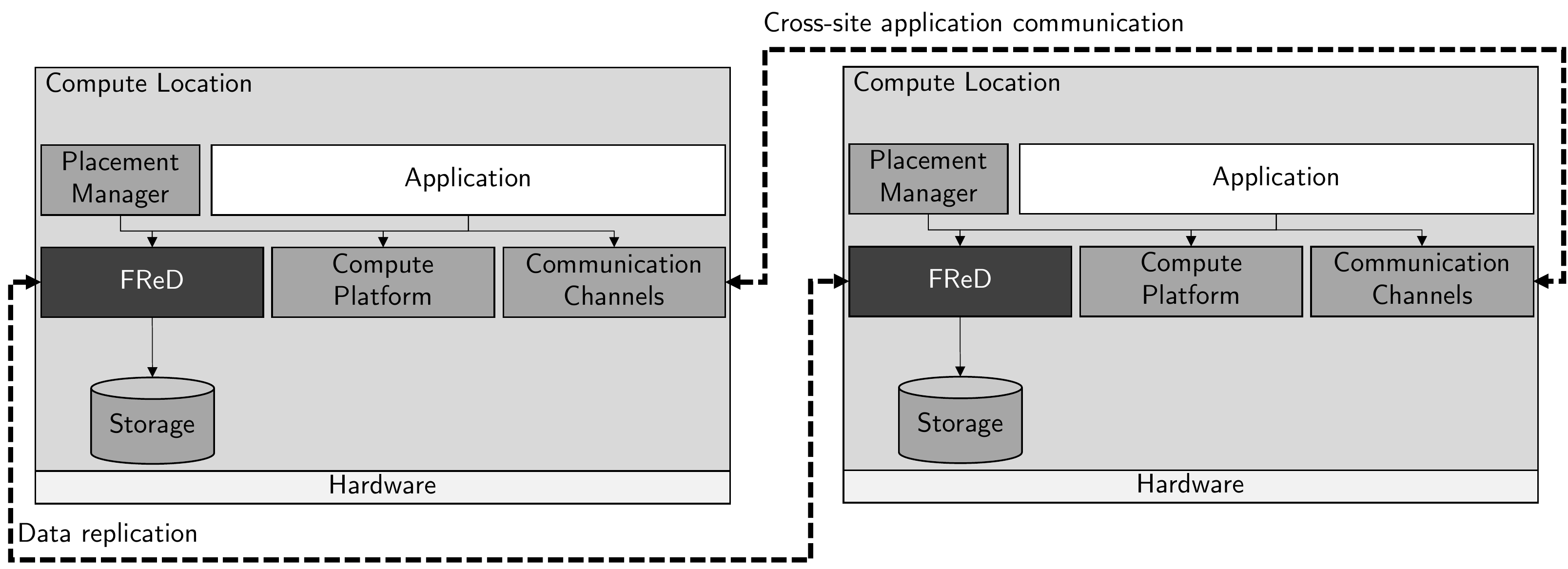}
    \caption{\textbf{The Integration of FReD in the Fog Application Stack:} FReD replicates data between compute locations, with a local storage backend at each location. A placement manager, which can be part of the application or an independent component, manages data replication policies. The application code runs on a compute platform and uses communication channels to distribute messages across sites.}
    \label{fig:landscape}
\end{figure}

Data management in the fog is only a part of the necessary supporting technologies for fog applications.
As shown in \cref{fig:landscape}, FReD should be run alongside compute and messaging platforms that manage service execution and message distribution, respectively.
Application code that itself is mostly stateless or only contains session-specific state, such as those designed for FaaS or other serverless systems, can interact with a local FReD instance for data access.
Under the hood, FReD uses a local database system or the file system for data persistence and replicates data to other compute locations.

FReD itself does not decide on replica placement.
Instead, FReD exposes application-controlled replica placement as an easy-to-use abstraction to applications.
A \textit{placement manager} that is part of the application code or an independent (partially) distributed and decentralized component uses this abstraction to manage replica placement and replication strategies.
We have outlined approaches for independent placement management in previous work~\cite{paper_pfandzelter_predictive_replica_placement,paper_bellmann_predictive_replication_markov}.
Likewise, the placement manager would control placement of application components across different compute locations -- it could possibly even be integrated into the Kubernetes scheduler.

\subsection{Prototype Implementation}
\label{sec:design:prototype}

To show the feasibility of our design and to evaluate FReD, we also present a proof-of-concept implementation.
The main node software and our client library use the \emph{Go}\footnote{\url{https://go.dev/}} programming language, but clients can use a variety of languages as the API uses \emph{gRPC}\footnote{\url{https://grpc.io/}}.
Internal communication between FReD nodes also uses gRPC.
If a node comprises several machines, requests are sharded based on keygroup through a common FReD proxy.
We have implemented storage adapters for in-memory storage, storage on disk, and AWS DynamoDB, and new storage backends can be easily added through the \texttt{Storage} interface.
The naming service is based on the \emph{etcd}\footnote{\url{https://etcd.io}} distributed key-value store.
External trigger services are integrated using a common gRPC interface: FReD can use any endpoints that implement this interface, hence there are no limits imposed on possible deployment models or programming languages.

%% file: sections/5_evaluation.tex
\section{Evaluation}
\label{sec:evaluation}

The evaluation comprises two parts.
First, in \cref{subsec:eval_consistency}, we show that FReD's distributes data consistently across nodes.
Second, in \cref{subsec:eval_use-cases}, we showcase how simple it is to build a typical fog computing application with FReD based on three example use cases.
We make all source code available on as open-source software\footnote{\url{https://github.com/OpenFogStack/FReD-scenarios}}.

\subsection{Consistency Experiments}
\label{subsec:eval_consistency}

We first evaluate data consistency within FReD using a simple distributed application adapted from Bermbach et al.~\cite{paper_bermbach_middleware_for_causal_consistency}.
In this application, online forum users concurrently post articles to one of ten conversation threads.
After randomly selecting a conversation, the client first reads the entire thread, then adds a new post to the thread and writes this new thread back into the database.
If a write conflict occurs, i.e., two clients try to add a new article at the same time, it should be resolved by adding both new posts rather than dropping one.
The database is a keygroup replicated across ten FReD nodes arranged in a fully meshed topology, where we inject 50ms RTT latency between any two nodes and between nodes and clients.
Each client randomly selects one of the FReD nodes and periodically (10\% probability) changes its chosen node.
This application is not representative of typical fog applications, but it serves as a ``worst-case'' example for concurrent data access that may cause consistency issues that would affect all kinds of distributed applications.

We vary the number of concurrent clients between 1 and 30.
Each client in our test completes 1,000 read and 1,000 write operations.
We measure if and how MRC or RYWC guarantees are violated, and for which operations FReD prevents dropping concurrent updates.
We repeat this experiment three times.
To compare the impact of the client library, we repeat this experiments once with all clients using the library and once without the library.

\begin{figure*}
    \centering
    \begin{subfigure}{0.66\textwidth}
        \centering
        \includegraphics[width=\linewidth]{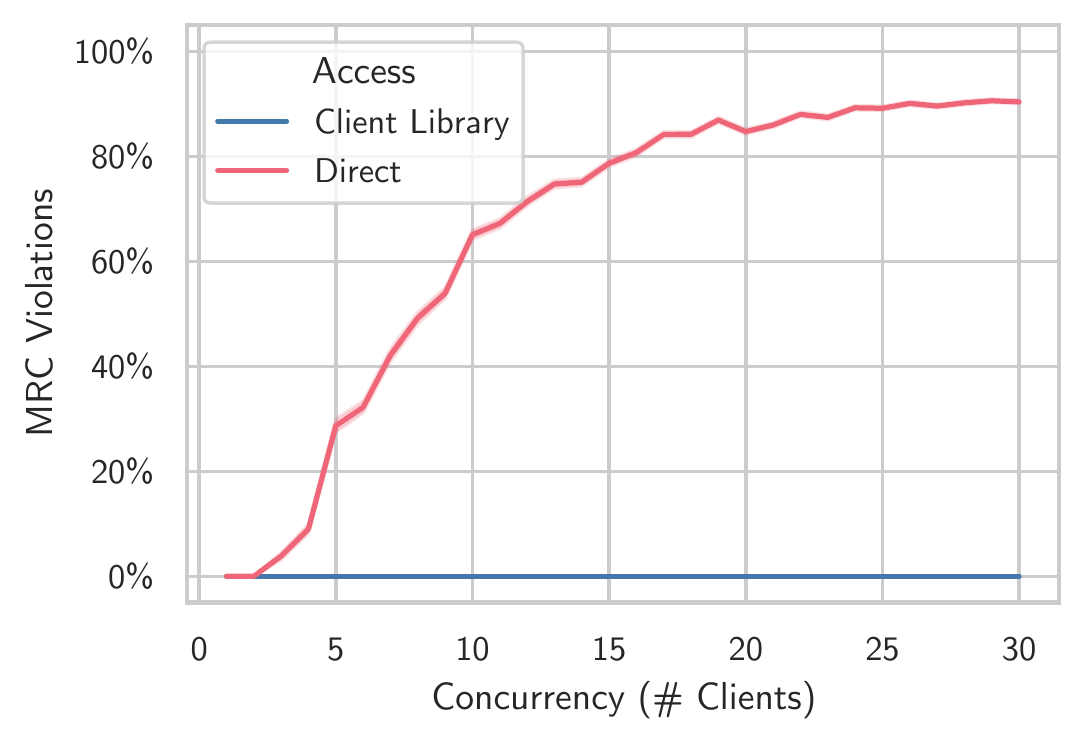}
        \caption{Read operations that result in MRC violations}
        \label{fig:consistency_eval:errors:mrc}
    \end{subfigure}
    \vfill
    \begin{subfigure}{0.66\textwidth}
        \centering
        \includegraphics[width=\linewidth]{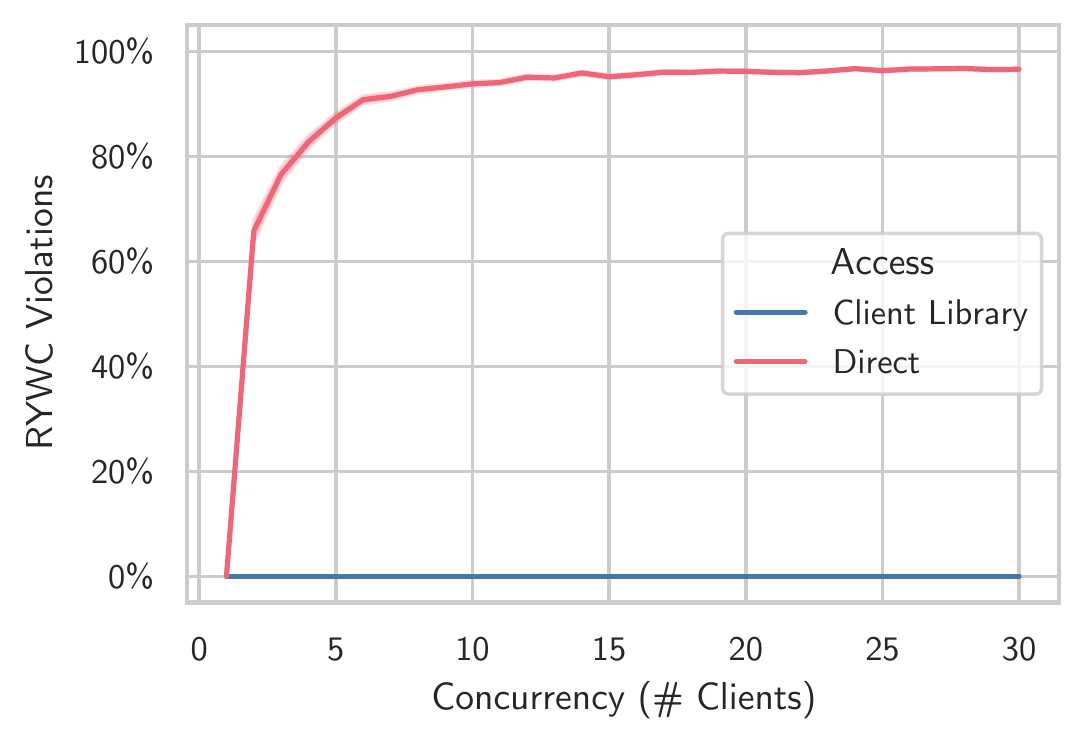}
        \caption{Read operations that result in RYWC violations}
        \label{fig:consistency_eval:errors:rywc}
    \end{subfigure}

    \caption{
        \textbf{Consistency Errors at Different Levels of Concurrency:}
        Using the client library guarantees both MRC (\cref{fig:consistency_eval:errors:mrc}) and RYWC (\cref{fig:consistency_eval:errors:rywc}) violations.
        With direct access from clients to FReD, where clients ignore the versioning information stored in FReD, the rate of consistency guarantee violations depends on the access concurrency.
        While few concurrent clients lead to few MRC violations, 90.4\% of read operations violate MRC consistency with 30 concurrent clients.
        RYWC is violated at an even greater rate, with 87.3\% violations with five concurrent clients and 96.6\% violations with 30 concurrent clients.
    }
    \label{fig:consistency_eval:errors}
\end{figure*}

\begin{figure*}
    \centering
    \begin{subfigure}{0.495\textwidth}
        \centering
        \includegraphics[width=\linewidth]{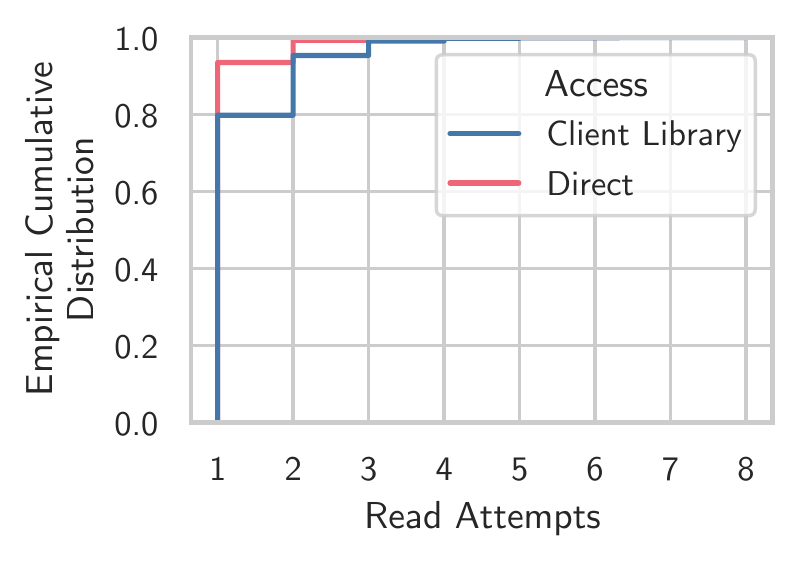}
        \caption{Attempted Reads per Operation}
        \label{fig:consistency_eval:attempts:read}
    \end{subfigure}
    \hfill
    \begin{subfigure}{0.495\textwidth}
        \centering
        \includegraphics[width=\linewidth]{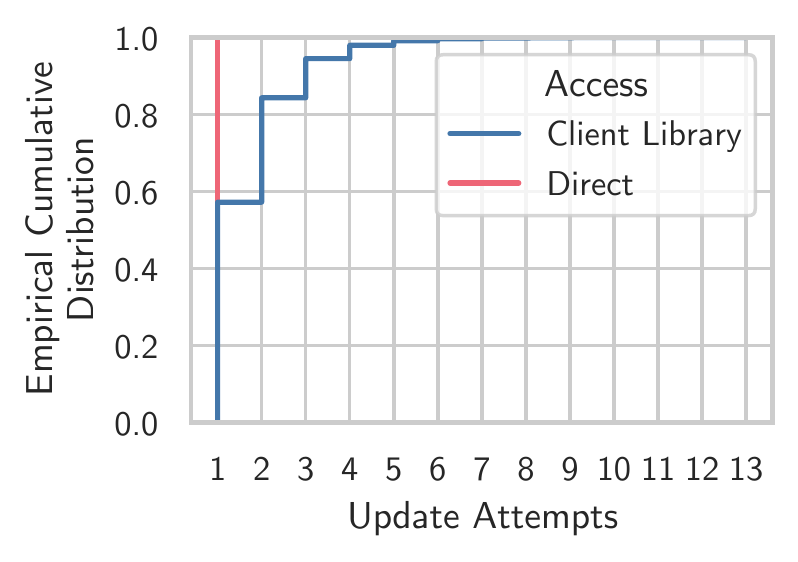}
        \caption{Attempted Updates per Operation}
        \label{fig:consistency_eval:attempts:update}
    \end{subfigure}
    \vfill
    \begin{subfigure}{0.495\textwidth}
        \centering
        \includegraphics[width=\linewidth]{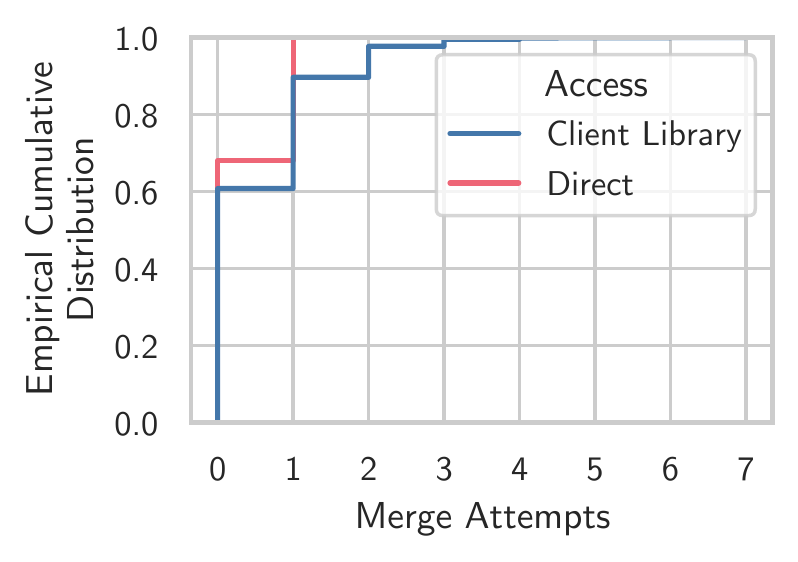}
        \caption{Attempted Merges per Operation}
        \label{fig:consistency_eval:attempts:merge}
    \end{subfigure}

    \caption{
        \textbf{Repeat Attempts with Erroneous Operations with 30 Concurrent Clients:}
        When the client library detects an operation that would violate a consistency guarantee, it will return an error to the client and prompt it to attempt the operation again.
        We thus observe increased attempts for read (\cref{fig:consistency_eval:attempts:read}), update (\cref{fig:consistency_eval:attempts:update}), and merge (\cref{fig:consistency_eval:attempts:merge}) operations.
        Without the client library, read operations can still fail if the requested data item is not yet replicated at the time of access.
        Update operations do not fail, even if they overwrite newer data.
        As FReD captures causality internally, merge operations may still be necessary without the client library when two updates occur concurrently.
    }
    \label{fig:consistency_eval:attempts}
\end{figure*}

We show the number of MRC and RYWC violations as a function of concurrency in \cref{fig:consistency_eval:errors}.
As expected, we observe no violations of MRC or RYWC when using the client library.
With direct access, however, we see increasing rates of violations as the concurrency increases, with 90.4\% MRC violations and 96.6\% RYWC violations at 30 concurrent clients.

The cost for this is an increased number of rejected read, update, and merge operations as shown in \cref{fig:consistency_eval:attempts}.
With the client library, 20.2\% read operations had to be repeated at least once as the library had detected outdated data.
Without the library, 6.5\% of reads had to be repeated, which could be caused by slow replication between FReD nodes given the network latency.

Similarly, the library would also support in rejecting update operations if data was to be overwritten that had not yet been read by the client.
As a result, 42.7\% update operations were rejected on the first try and had to be repeated.
Without the client library, all update operations were accepted by FReD, which comes at the cost of consistency guarantees.

Further, we see that clients had to perform merge operations, which indicates that at several points during our test, concurrent versions of the same data item existed.
These concurrent versions are caused by two or more clients updating the same value at different nodes concurrently, which can happen both with and without the client library.
Since no order can be inferred by FReD from such updates, both versions are stored and clients can choose to merge them.
Using a different strategy, such as \emph{Last Writer Wins}, those updates would be lost on propagation among nodes.

\subsection{Use Cases} \label{subsec:eval_use-cases}

In this section, we showcase how simple it is to build a typical fog computing application with FReD based on three example use cases.
We have implemented the application components for each use case in a different programming language.
As a metric for simplicity, we report the lines of code as counted by \texttt{scc}\footnote{\url{https://github.com/boyter/scc}}.

\subsubsection{Use Case 1: Mobile App}
\label{subsec:eval_usecases_app}

\begin{figure}
    \centering
    \includegraphics[width=\textwidth]{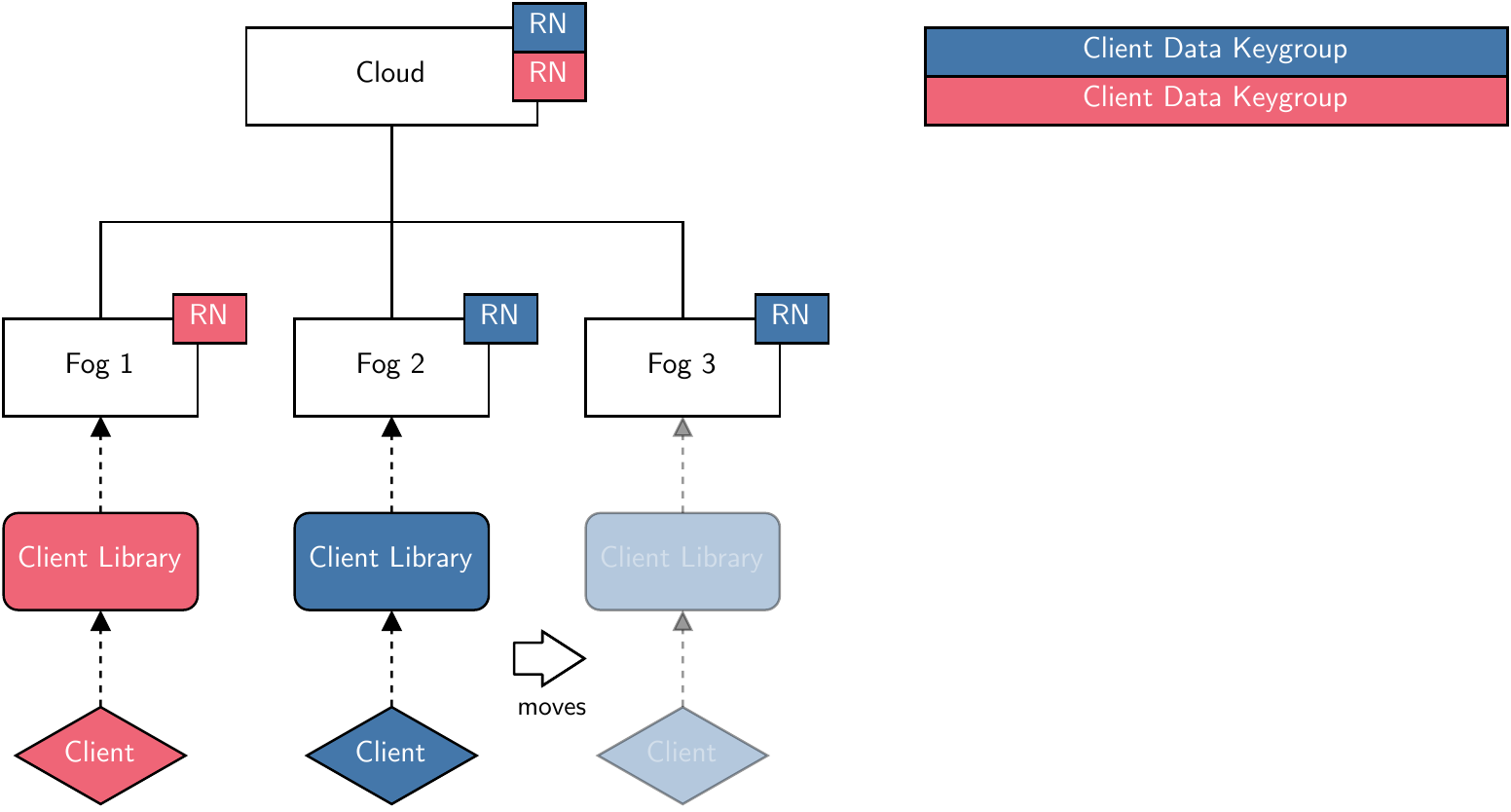}
    \caption{
        \textbf{Mobile App System Architecture:}
        Clients move through the fog and connect to different fog nodes.
        Each client has a dedicated client data keygroup that is replicated on the client itself and on the nearest fog node.
        Further, a cloud node is used to persist data.
    }
    \label{fig:eval_app}
\end{figure}

In this scenario, a mobile device such as a smartphone or tablet runs an app that uses the cloud for persistence.
To speed up data retrieval, data is also stored at nearby edge nodes.
Depending on movement of the user through the physical world, the fog node used for data replication may change.
Crucially, this allows the app to access data with low network overhead even in case of network partitions and interruptions between edge and cloud, e.g., when routing errors or latency spikes occur.
Furthermore, through the use of the local client library on the client device, FReD can ensure client-centric consistency guarantees.
We illustrate this scenario in \cref{fig:eval_app}.

FReD makes it possible to migrate data from one edge node to another by simply specifying where the data is needed rather than having to handle the complexity of actual data replication.
We have implemented this complete application in only \textbf{223 lines of Go code}.

Each client sets up one client data keygroup with at least two replica nodes: a nearby edge node and a cloud node.
The purpose of this keygroup is to store the client's data nearby for low latency access while also making sure that a backup copy is readily available in the cloud, e.g., for access by a further backend service.
When a client notices that it moves towards another edge node, it can add this fog node to its keygroup to preemptively initiate data replication.
Removing the previous fog node from the keygroup ensures that no unnecessary data copies exist.

\subsubsection{Use Case 2: Remote Research Stations}
\label{subsec:eval_usecases_research}

\begin{figure}
    \centering
    \includegraphics[width=\textwidth]{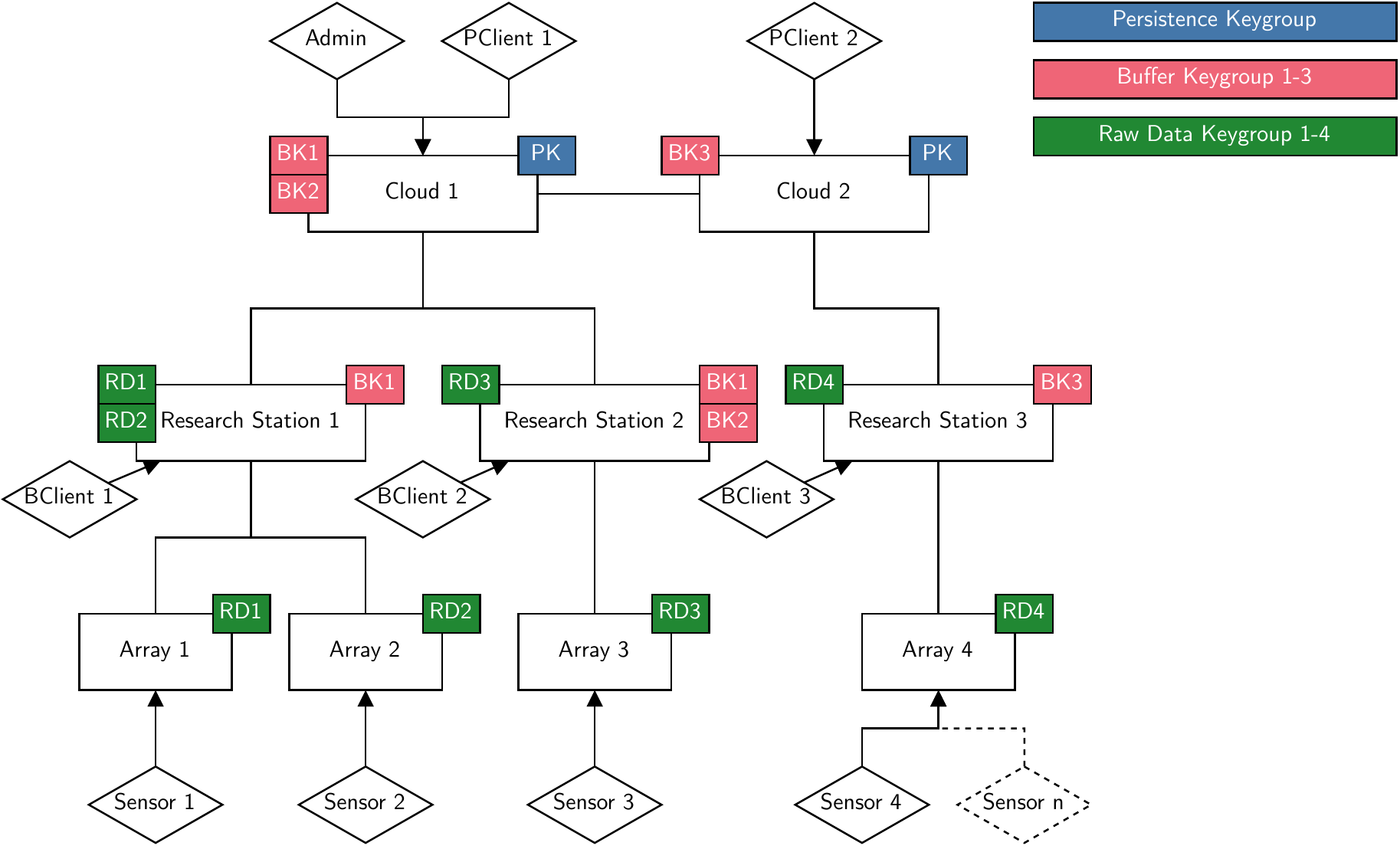}
    \caption{
        \textbf{Remote Research Station System Architecture:}
        Multiple remote sensor arrays collect data from various sensors.
        Each of the sensor arrays is attached to a specific research station, where raw data is replicated.
        A buffer keygroup is used to replicate data to the cloud, where it is persisted in a dedicated keygroup.
        Clients can access data both locally in the research stations, or remotely in the cloud.
    }
    \label{fig:eval_rrs}
\end{figure}

In this use case, off-site research stations, e.g., in the polar region, manage multiple sensor arrays that collect data from various sensors~(\cref{fig:eval_rrs}).
This data should be available in different forms for different amounts of time at various locations.
FReD makes it possible for the software engineers to simply define these constraints without having to worry about actually implementing this behavior.

For this purpose, three kinds of keygroups need to be set up.
To showcase that the keygroup setup can be done at a centralized location, we use an admin client that only communicates with \texttt{Cloud 1}.
The keygroup configuration is then distributed to all other nodes via FReD.
We managed to implement the complete scenario with only \textbf{690 lines of Java code}.

\paragraph{Raw Data Keygroup}
There is one raw data keygroup for each array.
Besides the array, this keygroup also comprises the closest research station.
Sensor clients write their measurements to the keygroup; FReD then replicates the data to the research station when an uplink is available.
A buffer client (\texttt{bclient}) that is connected to the research station regularly retrieves data from the keygroup for processing and aggregation.
After such operations, the \texttt{bclient} also removes the processed items from the keygroup at the research station.
Consequently, the data will also be removed from the array automatically by FReD, which frees up storage.

\paragraph{Buffer Keygroup}
There is one buffer keygroup for each research station.
Besides the research station, this keygroup also comprises the closest cloud node.
For increased availability, it is also possible to add another nearby research station to this keygroup as done for BK1.
Buffer clients use the buffer keygroup to store their processing and aggregation results.
Furthermore, a persistence client (\texttt{pclient}) that is connected to the cloud node regularly persist data from the keygroup for long term storage.
After such operations, the \texttt{pclient} also removes the persisted items from the keygroup at the cloud.
Consequently, the data will also be removed from the research station(s) automatically by FReD, which frees up storage.

\paragraph{Persistence Keygroup}
There is one persistence keygroup that is shared among all cloud nodes.
Persistence clients use this keygroup to store the data they retrieve from the individual buffer keygroups.
Since FReD replicates the data across all cloud nodes, data is secured against natural disasters.
Furthermore, other kinds of clients can access it at the nearest cloud region.

\subsubsection{Use Case 3: Smart Metering}
\label{subsec:eval_usecases_smartmeter}

\begin{figure}
    \centering
    \includegraphics[width=\textwidth]{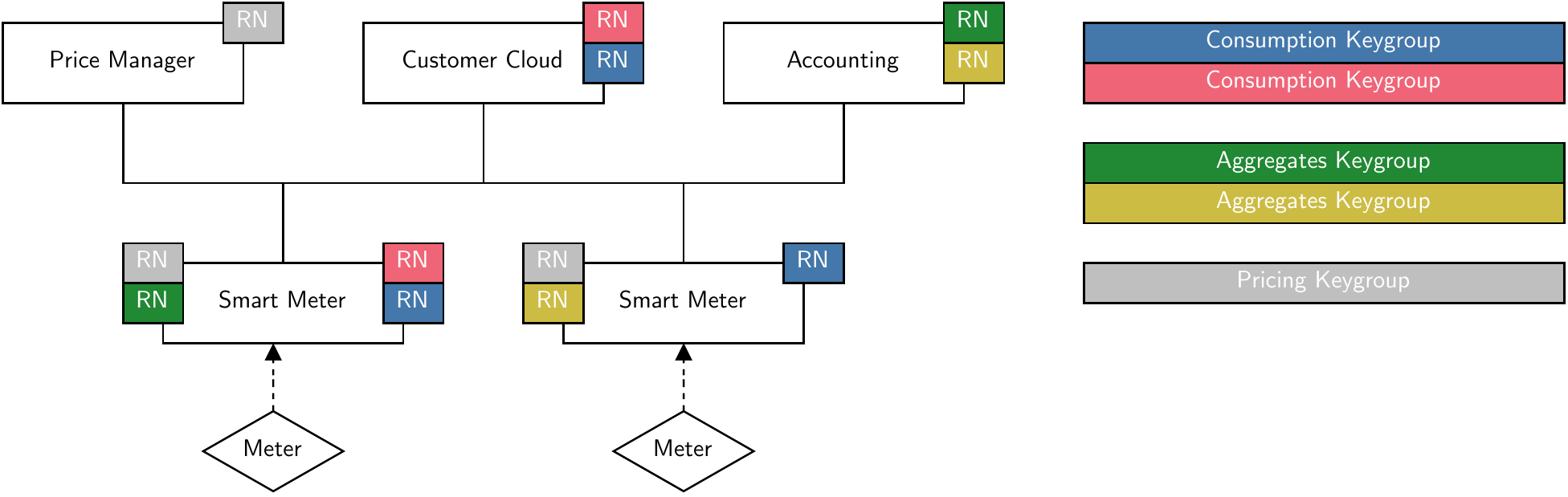}
    \caption{
        \textbf{Smart Meter System Architecture:}
        A smart meter makes local decisions based on data collected locally and sourced from different data sources.
        It receives pricing data by replicating data from the Price Manager, and replicates aggregate data to an Accounting component.
        Further, data may be replicated in a customer cloud or on other smart meters.
    }
    \label{fig:eval_meter}
\end{figure}

In this scenario, an electricity supplier offers a renewable energy contract that considers the current production of renewables.
Depending on the current production, the end-user price during a specific time frame is either \emph{high} or \emph{low}.
For privacy reasons, detailed consumption statistics should not be transmitted to the electricity supplier but only be available locally via a smart meter dashboard.
Thus, only daily aggregates that describe how much energy was consumed for the high and low prices are transmitted.
Furthermore, the smart meters need local access to information that describe whether a given time period had a high or low price.
FReD makes it possible to simply define where the data is needed instead of having to implement the for this needed interaction between the various components.
For this purpose, we simply set up three kinds of keygroups, as shown in \cref{fig:eval_meter}.
Our complete implementation of this scenario comprises only \textbf{547 lines of Go code}.

\paragraph{Consumption Keygroup}

Each household sets up one consumption keygroup which stores finely-grained energy consumption data that can be visualized and explored via a local dashboard.
We can use an append-only keygroup here, as time series data will not be modified afterwards.
Per default, the only replica node in this keygroup is the smart meter of the household.
Furthermore, owners can decide to optionally add a neighbor's house or a dedicated customer cloud to the keygroup for increased data availability to prevent paying the high price for data gaps.
To limit storage requirements, we simply set up this second replica with a time-to-live of 24 hours, i.e., data is automatically removed from this backup replica after one day.

\paragraph{Pricing Keygroup}

A price manager uses the pricing keygroup to store data on energy pricing (high or low) for each period of time.
Besides this price manager, all smart meters are added as replica nodes to this keygroup so that data is available in each household.

\paragraph{Aggregates Keygroup}

At the end of each day, each smart meter uses the data from the consumption keygroup as well as the pricing keygroup to compute how much energy was consumed for the high and low price.
We implement this with a trigger set to the consumption keygroups on each of the smart meters:
A small service calculates the consumed energy, taking into account pricing data, and writes that average back into FReD.
The accounting node can access this aggregated data by replicating it locally.

\subsection{Requirements Evaluation}

As a final part of our evaluation, we again consider the requirements for a fog data management system as outlined in \cref{subsec:requirements}.

\paragraph{RQ1: Common Data Access Interface}
Applications can interface with FReD using a standard key-value CRUD API.
Crucially, FReD maintains this API regardless of which underlying datastore is used, from in-memory storage on constrained edge nodes to full cloud storage.
We can thus state that FReD provides a common data access interface across the fog environment.

\paragraph{RQ2: Adapt to Infrastructure Heterogeneity}
Further, the storage a\-dap\-ters in FReD also let it adapt to infrastructure heterogeneity.
Any storage interface can provide such an adapter without implementation efforts by an application and without changes to FReD itself.
As the fog ecosystem evolves, applications that use FReD can continue to evolve alongside without maintenance efforts by application developers.

\paragraph{RQ3: Transparent Geo-Distribution}
The keygroup abstraction of FReD allows applications to specify where data is replicated and where it is not replicated.
Through the naming service, an application can check the geographical distribution of a keygroup at any time.
FReD thus also provides transparency on data distribution.

\paragraph{RQ4: Low-Hop Data Access}
The aforementioned transparent geo-dis\-tri\-bu\-tion combined with application-controlled replica placement also allows application services to replicate data close to clients, as we have also shown in the Mobile App use case in \cref{subsec:eval_usecases_app}.
Additionally, clients can make requests directly to FReD nodes without accessing a central coordinator for each request.
This means that the hot path of data access is limited only by network distance between a client and a data replica.

\paragraph{RQ5: Integrate Aggregation and Filtering}
The powerful trigger node abstraction in FReD allows integrating reactive services, such as filters and aggregation, at any FReD node and in any programming language.
As shown in \cref{subsec:eval_usecases_research}, this also facilitates higher-order data-based applications such as processing pipelines on top of FReD.

\paragraph{RQ6: Control Data Movement}
In FReD, applications have full control over data movement.
Keygroup data is only replicated to data replica nodes, which are only set by the application that uses this data.
For example, a FReD deployment that includes a cloud node will only replicate data to this cloud if explicitly instructed to do so.
Although FReD includes a naming service, this only stores some metadata and no application data.
Further, trigger nodes are attached directly to FReD nodes as well, which means that data movement for processing can also be controlled directly.
As we have shown in \cref{subsec:eval_usecases_smartmeter}, FReD can thus also be used to build privacy-conscious applications.

\paragraph{RQ7: Topology-Agnostic}
FReD does not follow a specific fog topology, as FReD nodes are considered equal regardless of the underlying infrastructure.
Although hierarchical topologies can also be reflected in FReD, a FReD deployment can also manage just edge nodes or just multiple clouds.
FReD can also manage data replication on changing topologies, as it uses direct data replication between nodes.

\paragraph{RQ8: Application-Agnostic}
FReD uses a key-value interface on any node, making it agnostic regarding the application it manages data replication.
Further, the keygroup abstraction allows arbitrary data replication policies and can thus enable a wide variety of fog applications.

\paragraph{RQ9: Manage Data Consistency and Availability}
As part of FReD, we have developed a client-side library that manages client-centric data consistency if required.
Applications can choose to use this client library to achieve a specific consistency guarantee, or they can use FReD directly to achieve higher availability in case of network partitions.

%% file: sections/6_discussion.tex
\section{Discussion}
\label{sec:discussion}

In our evaluation, we have shown that FReD fulfills the requirements for a fog data management system we have set.
In this section, we discuss threats to validity as well as limitations of our work.
Additionally, we also derive future work on FReD and fog data management in general.

\paragraph*{Burden of Data Replication Management for Developers}

While FReD can manage data replication for applications, it can be unclear if this provides a benefit for developers.
This depends mainly on the complexity of data replication, e.g., how many data replica locations must be managed.
For an application that only transfers data from a single sensor to a cloud database, for example, this complexity may be limited and FReD may be too large a dependency to consider.
In our use case evaluation in \cref{subsec:eval_use-cases}, however, we have shown how even small applications can benefit from FReD, which makes handling consistency, replication, and integrating processing easier.
We project that in typical fog applications, complexity of data movement is even greater~\cite{paper_pfandzelter_zero2fog}.

\paragraph*{Limits of the Key-Value Interface}

The keygroups in FReD encapsulate key-value data items.
While we assume that most fog applications will require data that fits this format, such as logs of IoT sensor data, application state, or even encoded images,
there may be applications that are limited by such an interface.
Although our implementation can be extended with support for structured data, i.e., columns, other formats do not fit the abstraction of per-item replication.
For blob data, such as large machine learning models, replication can be slow and resource-intensive if updated frequently.
Streams of data, such as live video streams and other data that cannot be divided into smaller parts, can also not benefit from per-item replication.

\paragraph*{Feasibility of Platform-Assisted Replica Placement}

In FReD, applications control replica placement directly, which provides benefits for access latency and privacy.
Although FReD eases fog application management, application developers still have to develop their own strategies for replica placement.
In future work, we hope to develop mechanisms to have the fog data management platform assist applications in placing replicas to help achieve certain QoS goals.
In previous work~\cite{paper_pfandzelter_predictive_replica_placement,paper_bellmann_predictive_replication_markov}, we have shown how this could be achieved by predicting future data access locations to move data replicas to an optimal replica location.

\paragraph*{Relevance of Fog Data Management in Practice}

Current trends in fog computing are driven mainly by research, and we have yet to see broad adoption of fog computing in industry.
This limits what is currently known about the constraints at play in real fog environments and requirements for fog data management approaches may have to be adapted as more information is known.
Note that this is not a limitation of this work specifically but could affect fog computing research in general.
In fact, FReD may have application beyond the fog, e.g., to manage data replications in federated or hybrid clouds~\cite{paper_kurze_cloud_federation}.

%% file: sections/7_conclusion.tex
\section{Conclusion \& Future Work}
\label{sec:conclusion}

In this paper, we have presented FReD, a data replication middleware for the fog that takes the burden of data replication, distribution, and mobility off of fog application developers.
With FReD, applications simply specify \emph{keygroups}, logical groups of key-value items, and their replica locations across the fog network.
FReD manages data replication across any fog topology and even provides mechanisms for data consistency with a client library and for building data processing applications with trigger nodes.
Further, with the integration of \emph{triggers}, developers can use the data distribution capabilities of FReD in fog application workflows and execute processing services in response to data updates.
We have evaluated FReD using three example applications and extensively compared it to the state of the art in distributed data management in the cloud and fog.

Our survey of related work has revealed a research gap in fog data management systems that abstract the complexity of fog computing to a high degree while being applicable to any fog application.
FReD fills this gap in the fog application stack.
In future work, we plan to extend this approach with mechanisms for managing other kinds of data, such as video streams.
Further, we plan on building on top of the FReD abstractions to evaluate approaches for assisting applications in replica placement, e.g., using client movement prediction.
Finally, we plan on exploring to what extent the abstractions developed with FReD are applicable to compute service and messaging middlewares for the fog.